\documentclass[aps,pra,letterpaper,superscriptaddress,10pt,notitlepage,twocolumn]{revtex4-1}
\usepackage{amsmath,amssymb}
\usepackage{graphicx,epstopdf,ulem}
\usepackage[pdftex,dvipsnames,usenames]{xcolor}
\usepackage[colorlinks=true,urlcolor=blue,citecolor=blue,linkcolor=blue]{hyperref}

\begin{document}

\title{Benchmarking of Gaussian boson sampling using two-point correlators}

\author{D. S. Phillips}
	\email{david.phillips@physics.ox.ac.uk}
	\affiliation{Clarendon Laboratory, University of Oxford, Parks Road, Oxford OX1 3PU, United Kingdom}

\author{M. Walschaers}
	\affiliation{Laboratoire Kastler Brossel, Sorbonne Universit\'e, CNRS, ENS-PSL Research University, Coll\`ege de France; 4 place Jussieu, F-75252 Paris, France}

\author{J. J. Renema}
	\affiliation{Complex Photonic Systems, Faculty of Science and Technology, University of Twente, P.O. Box 217, NL-7500 AE  Enschede, The Netherlands}

\author{I. A. Walmsley}
	\affiliation{Clarendon Laboratory, University of Oxford, Parks Road, Oxford OX1 3PU, United Kingdom}

\author{N. Treps}
	\affiliation{Laboratoire Kastler Brossel, Sorbonne Universit\'e, CNRS, ENS-PSL Research University, Coll\`ege de France; 4 place Jussieu, F-75252 Paris, France}

\author{J. Sperling}
	\affiliation{Clarendon Laboratory, University of Oxford, Parks Road, Oxford OX1 3PU, United Kingdom}

\date{\today}

\begin{abstract}
	Gaussian boson sampling is a promising scheme for demonstrating a quantum computational advantage using photonic states that are accessible in a laboratory and, thus, offer scalable sources of quantum light.
	In this contribution, we study two-point photon-number correlation functions to gain insight into the interference of Gaussian states in optical networks.
	We investigate the characteristic features of statistical signatures which enable us to distinguish classical from quantum interference.
	In contrast to the typical implementation of boson sampling, we find additional contributions to the correlators under study which stem from the phase dependence of Gaussian states and which are not observable when Fock states interfere.
	Using the first three moments, we formulate the tools required to experimentally observe signatures of quantum interference of Gaussian states using two outputs only.
	By considering the current architectural limitations in realistic experiments, we further show that a statistically significant discrimination between quantum and classical interference is possible even in the presence of loss, noise, and a finite photon-number resolution.
	Therefore, we formulate and apply a theoretical framework to benchmark the quantum features of Gaussian boson sampling under realistic conditions.
\end{abstract}

\maketitle

\section{Introduction}

	In their seminal work \cite{HBT56}, Hanbury Brown and Twiss analyzed two-point correlators to improve the apparent angular size estimation of distant stars.
	On the quantum level, two-point correlations render it possible to experimentally uncover nonclassical properties of light, e.g., photon antibunching \cite{KDM77}.
	Nowadays, general quantum correlations form the foundation of quantum information and communication science \cite{NC10,LB99,KLM01}.
	For example, continuous-variable entanglement offers a robust resource for quantum protocols when optical modes propagate through the turbulent atmosphere \cite{BSSV16,SEFLBHSU17}.
	While quantum correlations enable us to perform certain tasks, such as quantum teleportation \cite{BBCJPW93,Retal17}, the problem of whether or not there is a true advantage of quantum protocols over classical information processing is still debated as quantum correlations can be significantly diminished in the presence of imperfections and require error correction (see Ref. \cite{B18} for a recent popular discussion).
	For these reasons, the application-oriented study of realistic quantum correlations is a timely problem of fundamental importance for the development of quantum technologies.

	A promising scheme to demonstrate the advantage of quantum computers over classical computers is boson sampling \cite{AA13}.
	This scheme comprises sending indistinguishable photons into a multiport interferometer, for example, made up of variable beam splitters and phase shifters, and measuring the photon-number distribution from the output.
	The multiport interferometer implements a unitary transformation of the bosonic modes that in turn yields a highly entangled output state.
	Calculating the output probability for a given configuration is related to calculating the permanent of the unitary transformation matrix \cite{S04}, which is a computationally hard problem as it scales exponentially with the size of the system \cite{V79}.
	Therefore, building a device which could sample from the output of an interferometer faster than a classical computer could do would unambiguously demonstrate a quantum computational advantage.

	One problem with realizing the boson sampling protocol experimentally is that single photons are hard to generate efficiently.
	Common experimental methods rely on post-selection from the Gaussian states obtained from spontaneous parametric down conversion;
	however, post-selection does not scale favorably \cite{BFRDARW13,TDHNSW13,Setal13,CORBGSVMMS13}.
	To remedy this problem, scattershot boson sampling was introduced to effectively increase the number of down conversion sources.
	However, this scheme ultimately relies on a similar post-selection \cite{LLRRBR14,BSVFVLMBGCROS15}, thus making it prone to the same scaling problems.
	An alternative solution is to use more deterministic photon sources for boson sampling, such as quantum dots \cite{Wetal17,LBHGSLASW17}.
	Currently, though, the degree of indistinguishability between two different quantum dots is not high enough, and one must resort to a single dot with delay lines instead.
	The disadvantage of this approach can lead to an unfavorable scaling in time.

	A recent development in the field of boson sampling is to use Gaussian states as the inputs to the multiport interferometer.
	Gaussian states can be generated deterministically from spontaneous parametric down conversion sources.
	While calculating the probability of a given output photon-number configuration in the original boson sampling problem is related to calculating a matrix permanent, the Gaussian-boson-sampling equivalent is related to calculating the Hafnian of a matrix, which still lies in the same complexity class due to photon-number projection being a non-Gaussian measurement \cite{HKSBSJ17,KHSBSJ18}.
	It is important to note, however, that there are currently no rigorous hardness results for Gaussian boson sampling that tolerate (e.g., additive) errors.
	Still, some potential uses have been proposed for the protocol apart from proving a quantum computational advantage.
	These potential applications include the nontrivial simulation of complex molecular vibronic spectra \cite{AW12,HGPMA15,HY17,CREVLGNKHW17} (of which a proof-of-concept experiment has already been performed \cite{CREVLGNKHW17}), performing sophisticated calculations in graph theory \cite{ABR18,AB18}, and quantum machine learning \cite{SSMQWB18}.

	Beyond such practical considerations, even fundamental aspects of Gaussian boson sampling are still actively studied.
	The setup fits in the quest for achieving a true quantum advantage in the continuous-variable setting.
	It is well established that a non-Gaussian element is required to render a setup hard to simulate on a classical computer \cite{BSBN02,ME12,VWFE13,RRC16}.
	In scattershot and Gaussian boson sampling, the non-Gaussian features are introduced at the measurement stage through the use of photodetectors.
	Still, a quantum computational advantage has also been found in alternative scenarios with Gaussian detectors and non-Gaussian input states \cite{DMKDCMvF17,CC17,CDMvKF17,LRR17}.
	Furthermore, in standard and scattershot boson sampling, utilizing photodetectors and Fock states, one can identify the phenomenon of many-particle interference \cite{TTMB12,T14,TTSSdHNSW15}---a generalization of the Hong-Ou-Mandel effect \cite{HOM87}---as the source of the computational complexity.
	At present, it is unclear whether Gaussian boson sampling is just a manifestation of the same physical phenomenon, or whether there is additional physics to be uncovered in these setups which would lead to a different computational complexity condition.
	One approach to answering this question is to investigate how measurable signatures of many-particle interference change in Gaussian boson sampling by analyzing correlations.

	Signatures of many-particle interference also serve an important purpose as a tool for the benchmarking of boson sampling.
	The debate on how to validate a boson sampler started with the concern that it would be impossible to distinguish data from a boson sampling setup from data that were drawn from a uniform distribution.
	Thus, the first certification protocols aimed at making the distinction between these scenarios \cite{GKAE13,AA14,SVBBCFGMRMOGS14}.
	Even though this led to the development of several certification protocols \cite{AGKE15,BSVBCFRMOGS14}, the main focus in research on validation of boson sampling has shifted to hallmarks of many-boson interference \cite{CMSRIWRTOML14,TMBM14,Sh16,RDMWK16,VFICBSCBGOS18,WKUMTRB16,Getal18}.
	Furthermore, the alternative hypotheses for the origins of sampling data have gotten more physically motivated; most notably, one often probes the distinguishability of particles.

	In general, we can single out two approaches to benchmarking many-particle interference.
	On the one hand, one can construct highly symmetric unitary circuits (e.g., the Fourier interferometer \cite{TMBM14}) that manifest totally destructive interference, which are unique benchmarks of many-boson interference.
	On the other hand, one may instead use Haar-random circuits that are common in boson sampling and employ statistical analysis on the data (e.g., by studying two-point correlators \cite{WKUMTRB16}) to find statistical signatures of many-particle interferences. 
	
	Recent developments \cite{Dittel1,Dittel2} in the understanding of total destructive interference may provide a potential pathway for constructing a benchmark for Gaussian boson sampling.
	Nevertheless, the statistical signatures \cite{WKUMTRB16,WKB16,Getal18} are found by analyzing intensity correlations between output detectors, which can be calculated for arbitrary initial states.
	Therefore, this approach is a viable candidate for a benchmark of Gaussian boson sampling, a route which we extensively explore in this article.

	By benchmarking, we mean comparing the output correlations of quantum Gaussian input states (for instance, a squeezed vacuum) to a classical analog.
	The classical analog could be a coherent or thermal state---i.e., a classical state is erroneously prepared in a laboratory when the actually desired state is squeezed, where both of which are Gaussian states.
	By comparing the output correlations of the two states in the presence of experimental imperfections and limitations, one can determine the required accuracy to observe a meaningful difference between classical and nonclassical inputs.

	It is important to stress that our benchmarking scheme is an experimentally friendly way to distinguish different input states rather than being a sufficient condition to certify true Gaussian boson sampling.
	It could be used to complement a more robust verification scheme it can be applied directly to sampling data that are obtained from the Gaussian boson sampler.
	Indeed, the availability of an efficient and simple benchmark is an important step in the general endeavor of verification.
	The findings in Ref. \cite{AA13} already suggest that sampling from the output probability distribution can probably never be certified by a single verifier alone, thus emphasizing the need for several experimentally relevant benchmarking protocols.
	In addition to this, one of the findings of Ref. \cite{HKEG18} was that efficient, full certification of boson sampling that uses only the usual photon-number measurements is not possible.
	Therefore if this result extends to Gaussian boson sampling, then only benchmarking is possible using the measurements outlined in our scheme.

	In this paper, we investigate two-point correlation functions based on photon-number measurements to characterize boson sampling in continuous-variable systems, i.e., for general Gaussian states propagating in optical networks.
	Based on this method, we exploit the differences in the statistical signatures of the two-point correlation functions to discriminate Gaussian boson sampling with nonclassical (i.e., squeezed) from classical input states.
	Furthermore, it is shown that the phase dependence of squeezed states leads to additional contributions in correlators, unseen for rotationally invariant Fock states.
	Moreover, we complement our analysis by investigating the impact of a broad class of imperfections which can occur in realistic experimental realizations.

	The paper is organized as follows.
	In Sec. \ref{sec:CorrFunc}, we start by providing a general introduction to the two-point correlators, used in our benchmarking protocol.
	This is then supplemented by the framework of Gaussian quantum states in Sec. \ref{sec:Gaussian}, which is applied in Sec. \ref{sec:GaussianCorr} to find a closed expression for the relevant correlators.
	These expressions can then be averaged over the Haar measure by using techniques from random matrix theory to obtain the relevant statistical signatures, established in Refs. \cite{WKUMTRB16,WKB16}.
	In Sec. \ref{sec:Simulations}, we compare these statistical signatures to numerical simulations of Gaussian boson sampling, where we investigate the influence of squeezing on the correlators.
	Finally, in Sec. \ref{sec:ExpCons}, we apply the developed tools to carry out an in-depth analysis of experimental imperfections, relevant for future implementations.

\section{Correlation functions}\label{sec:CorrFunc}

	In statistical physics, a two-point correlator quantifies the correlation between two measured quantities.
	In general, correlators are second-order cumulants over multiple random variables, and higher orders can be generalized by the Ursell function \cite{S86}.
	These higher order correlators have a long history in quantum statistical mechanics as they characterize many-body states \cite{R65,BKR78,GVV89,NOS06} and are commonly referred to as truncated correlation functions.
	For two classical random variables, $X$ and $Y$, the two-point correlator $\mathbb C(X,Y)$ is commonly defined as
	\begin{equation}
		\label{eq:ClassCorr}
		\mathbb C(X,Y)
		=\mathbb{E}\left(XY\right)
		-\mathbb{E}\left(X\right)\mathbb{E}\left(Y\right),
	\end{equation}
	where $\mathbb{E}(\cdots)$ denotes the expectation value.

\begin{figure}[tb]
	\includegraphics[width=\columnwidth]{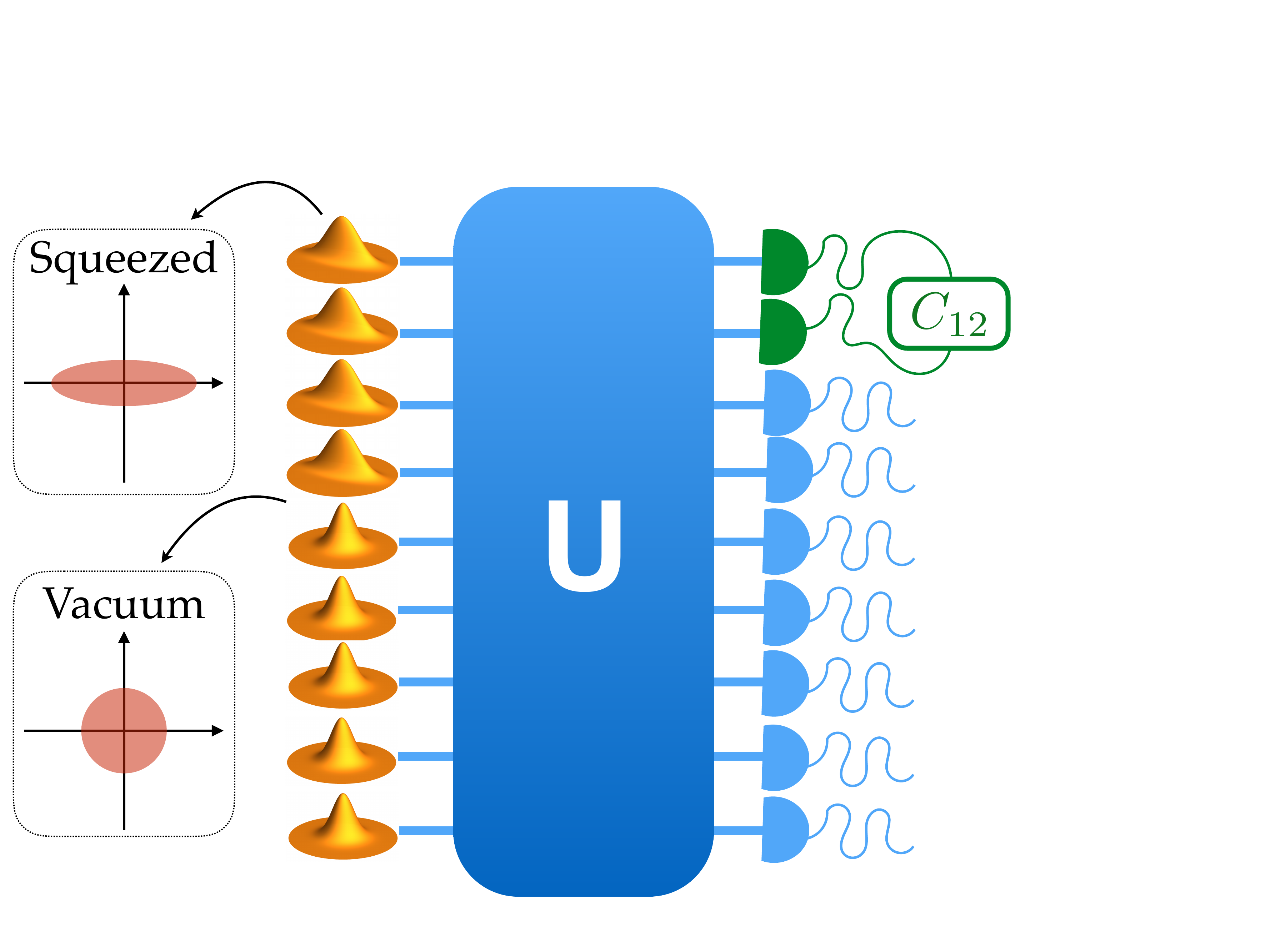}
	\caption{(Color online)
		Gaussian boson sampling scheme.
		In the depicted example, $N=4$ squeezed states are fed into an $M=9$ port interferometer, represented by a unitary $\boldsymbol U$.
		The photon-number correlation $C_{1,2}$ of two outputs is measured to apply the here-proposed benchmark.
	}\label{fig:setup}
\end{figure}

	Such correlations have been used to identify the statistical properties in the interest of benchmarking boson sampling with Fock states \cite{WKUMTRB16,Getal18}.
	Driven by the superior scaling of Gaussian boson sampling and the experimental feasibility to generate Gaussian states with down-conversion sources, we apply a similar analysis in order to benchmark boson sampling with phase-sensitive Gaussian quantum states against analogous classical states which can mimic some of the features of quantum Gaussian states.
	In Fig. \ref{fig:setup}, we outline the scenario under study in which a number of Gaussian input states are mixed in a unitary optical network.
	In particular, a two-point correlation measurement of two output ports is analyzed.

	As the conjectured hardness of Gaussian boson sampling arises from projecting the output states onto the photon-number basis, the two-point correlation function for Gaussian states is also considered in the number basis, in line with the analysis in Ref. \cite{WKUMTRB16}, rather than using the Gaussian quadrature correlations as obtained from balanced homodyne detection.
	The photon-number two-point correlation function $C_{j,k}$ on $2$ output modes $j$ and $k$ ($j,k\in\{1,\ldots,M\}$) is given by
	\begin{equation}
		\label{eq:TwoPointCorrelator}
		C_{j,k}
		=\langle \hat{n}_{j}\hat{n}_{k}\rangle
		-\langle \hat{n}_{j}\rangle \langle \hat{n}_{k}\rangle,
	\end{equation}
	where $\hat{n}_{j}$ is the $j$th photon-number operator and $\langle \cdots\rangle $ denotes the quantum-mechanical expectation value.
	This corresponds to the quantum-mechanical version of the classical expression in Eq. \eqref{eq:ClassCorr}.

	Two variants can be considered to implement the boson sampling procedure; cf. Ref. \cite{WKUMTRB16}.
	In the first scenario, one uses one fixed Haar-random unitary $\boldsymbol{U}$ to evolve the input state and then calculates $C_{j,k}$ for all output combinations $j<k$.
	The obtained set of correlators is then used as a data set for statistical tests, e.g., estimating moments of the correlators.
	In the second adaptation, one fixes the output ports (say $j=1$ and $k=2$, without loss of generality) and evolves the input state under many different Haar-random unitaries, i.e., unitary maps which are distributed according to the Haar measure.
	Here, the statistics is gathered by evaluating $C_{1,2}$ for each different realization of $\boldsymbol{U}$, which is closer to the analytical methods that are used to predict the statistical properties of the correlators.

	In the limit of a large number of modes, the correlations between the components of $\boldsymbol{U}$ are sufficiently small and both approaches become equivalent.
	However, practical reasons can make one implementation favorable over the other for a smaller number of modes $M$.
	For example, by fixing the unitary $\boldsymbol{U}$, the size of the dataset of correlators is automatically limited to $M(M-1)/2$, which might be insufficient statistical predictions.
	On the other hand, in many experimental setups (see, e.g., Ref. \cite{Getal18}), experimental constraints simply make it impossible to implement a large number of different realizations of $\boldsymbol{U}$.
	Nevertheless, in this article, we are able to explore the potential of Gaussian boson sampling with reconfigurable linear optics circuits and photon-number-resolving detectors on two output modes, which gives us the liberty to consider as many different realizations of $\boldsymbol{U}$ as required.

	To obtain statistical quantifiers of the resulting randomization process, the distribution of $C_{j,k}$ values can be analyzed.
	Note that the $j$ and $k$ indices are dropped in the following relations---meaning $C$ denotes an arbitrary element $C_{j,k}$---as all of the permutations are taken into account when the two-point correlator is averaged over many different Haar-random unitaries.
	\begin{subequations}
	\label{eq:RandomMatrixStatProp}
	For our purpose, the first characteristic is given by the normalized mean
	\begin{equation}
		\mathsf{NM}
		=\frac{
			\mathbb{E}_{U}\left(C\right)M^{2}
		}{N},
	\end{equation}
	the second one is the coefficient of variation
	\begin{equation}
		\mathsf{CV}
		=\frac{
			\sqrt{\mathbb{E}_{U}\left(C^{2}\right)-\mathbb{E}_{U}\left(C\right)^{2}}
		}{
			\mathbb{E}_{U}\left(C\right)
		},
	\end{equation}
	and the third quantity is the skewness
	\begin{equation}
		\mathsf{Sk}
		=\frac{
			\mathbb{E}_{U}\left(C^{3}\right)
			-3\mathbb{E}_{U}\left(C\right)\mathbb{E}_{U}\left(C^{2}\right)
			+2\mathbb{E}_{U}\left(C\right)^{3}
		}{\sqrt{{(
			\mathbb{E}_{U}\left(C^{2}\right)
			-\mathbb{E}_{U}\left(C\right)^{2}
		)}^{3}}}.
	\end{equation}
	\end{subequations}
	These three quantifiers correspond to the normalized first three moments of the distribution of $C_{j,k}$ for a fixed system averaged over many Haar-random unitaries, labeled as $\mathbb{E}_{U}\left(\cdots\right)$.
	A main objective of this work is to tell different families of Gaussian quantum states apart based on the values of $\mathsf{NM}$, $\mathsf{CV}$ and $\mathsf{Sk}$.

\section{Gaussian state formalism}\label{sec:Gaussian}

	In the quantum-optical description, each mode is represented through annihilation and creation operators, $\hat a_j$ and $\hat a_j^\dag$, respectively.
	We may collect the annihilation operator in the vector
	\begin{equation}
		\hat{\boldsymbol a}=(\hat a_1, \ldots, \hat a_M)^\mathrm{T}.
	\end{equation}
	The bosonic operators satisfy the commutation relation $[\hat a_j,\hat a_k^\dag]=\delta_{j,k}$, with $\delta$ denoting the Kronecker symbol.
	For each mode, we can write the photon-number operator as $\hat{n}_j=\hat a_j^\dag\hat a_j$.

	For the phase-space representation of optical fields, the quadrature representation is favorable, which is based on the operators
	\begin{equation}
		\hat{q}_j=\hat{a}_j+\hat{a}_j^\dag
		\quad\text{and}\quad
		\hat{p}_j=\frac{\hat{a}_j-\hat{a}_j^\dag}{i}.
	\end{equation}
	We then define the $2M$-dimensional vector of quadrature operators as
	\begin{equation}
		\label{eq:DispVec}
		\hat{\boldsymbol \xi}=(\hat{q}_1,\ldots,\hat q_M, \hat p_1,\ldots\hat p_M)^\mathrm{T}.
	\end{equation}
	Its expectation value corresponds to the location of the state in phase space, $\boldsymbol\xi_0=\langle\hat{\boldsymbol \xi}\rangle$.
	In addition, we get the $2M\times 2M$ covariance matrix $\boldsymbol V$ from the symmetric elements
	\begin{equation}
		\label{eq:CovMat}
		V_{j,k}=\frac{1}{2}\langle \Delta\hat\xi_{j}\Delta\hat\xi_{k}+\Delta\hat\xi_{k}\Delta\hat\xi_{j}\rangle,
	\end{equation}
	using the abbreviation $\Delta \hat x=\hat x-\langle\hat x\rangle$ for arbitrary operators $\hat x$.
	\begin{subequations}
	The covariance matrix also yields the covariances for the bosonic ladder operators,
	\begin{eqnarray}
		\label{eq:CovarRelA}
		\langle \Delta\hat a_j\Delta\hat a_k\rangle
		&{=}&\frac{
			V_{j,k}
			{+}iV_{j,k+M}
			{+}iV_{j+M,k}
			{-}V_{j+M,k+M}
		}{4},
		\\\notag
		\langle \Delta\hat a_j^\dag\Delta\hat a_k\rangle
		&{=}&\frac{
			V_{j,k}
			{+}iV_{j,k+M}
			{-}iV_{j+M,k}
			{+}V_{j+M,k+M}
		}{4}
		\\\label{eq:CovarRelB}
		&&{-}\frac{\delta_{j,k}}{2}.
	\end{eqnarray}
	Similarly, we identify complex displacements via
	\begin{equation}
		\label{eq:DisplRel}
		\langle \hat a_j\rangle=\frac{\xi_{0,j}+i\xi_{0,j+M}}{2}=\alpha_{0,j}.
	\end{equation}
	\end{subequations}
	Finally, an $M$-mode Gaussian state is equivalently given by a Wigner function which reads
	\begin{equation}
		W(\boldsymbol\xi)=
		\frac{\exp\left[
			-\frac{1}{2}
			\left(\boldsymbol\xi-\boldsymbol{\xi}_0\right)^\mathrm{T}
			\boldsymbol{V}^{-1}
			\left(\boldsymbol\xi-\boldsymbol{\xi}_0\right)
		\right]}{
			\sqrt{\left(2\pi\right)^{2M}\det\boldsymbol{V}}
		},
	\end{equation}
	where $\boldsymbol \xi$ includes the conjugate quadrature variables which define the $2M$-dimensional phase space.

	In the context of boson sampling, the annihilation operators of the input modes evolve under a unitary that describes the interferometer,
	\begin{equation}
		\hat{\boldsymbol{a}}\mapsto \boldsymbol{U}\hat{\boldsymbol{a}}.
	\end{equation}
	For the covariance matrix and the displacement vector, the transformation reads as follows:
	\begin{equation}
		\label{eq:VTrans}
		\boldsymbol{V}\mapsto\boldsymbol{O}\boldsymbol{V}\boldsymbol{O}^\mathrm{T}
		\quad\text{and}\quad
		\boldsymbol{\xi}_0\mapsto\boldsymbol{O}\boldsymbol{\xi}_0,
	\end{equation}
	where the orthogonal and symplectic transformation $\boldsymbol{O}$ is a $2M\times2M$ matrix defined by
	\begin{equation}
		\label{eq:UonV}
		\boldsymbol{O}
		=\left(\begin{array}{cc}
			\mathrm{Re}\left(\boldsymbol{U}\right) & -\mathrm{Im}\left(\boldsymbol{U}\right)
			\\
			\mathrm{Im}\left(\boldsymbol{U}\right) & \mathrm{Re}\left(\boldsymbol{U}\right)
		\end{array}\right),
	\end{equation}
	which is decomposed in separate blocks referring to the $q$ and $p$ components.

	Note that it is possible to calculate the probability of a given output photon-number configuration $P\left(\boldsymbol{n}\right)$, where $\boldsymbol{n}$ is an $M$-dimensional vector of output photon numbers in each mode from $\boldsymbol{V}$ and $\boldsymbol{\xi}_0$ alone.
	This can be done using multidimensional Hermite polynomials but involves rather complicated computations \cite{DMM94,KB01}.

	A premise of boson sampling is that the states entering the interferometer are uncorrelated.
	This means that all entries of the input covariance matrix which correlate different modes are zero.
	For this reason, we can characterize each single-mode input in terms of a $2\times 2$ covariance and a two-dimensional displacement vector.
	Further, a diagonalization can be achieved via a local unitary, which yields a single-mode covariance matrix of the form
	\begin{equation}
		\begin{pmatrix}
			\langle (\Delta\hat q_j)^2\rangle & 0
			\\
			0 & \langle (\Delta\hat p_j)^2\rangle
		\end{pmatrix}
		=\mathrm{diag}(v_{q,j},v_{p,j}).
	\end{equation}

	In such a diagonal form, the covariance matrix corresponds to a physical state if the uncertainty relation $v_{q,j}v_{p,j}\geqq1$ holds true.
	For $v_{q,j}=v_{p,j}=1$, we have a coherent or vacuum state, the latter for zero displacement.
	In the case that the variances are identical but larger than one, the input is a (displaced) thermal state.
	A (displaced) squeezed state is described when one of the variances is below the vacuum fluctuation, $v_{q,j}<1$ or $v_{p,j}<1$.
	For completeness, we could also have a classical state ($v_q,v_p\geq1$) which exhibits, however, an unequal noise distribution in the two quadratures ($v_p\neq v_q$).
	Such a state could simulate a squeezed vacuum state by having an asymmetric Wigner function that still has classical (i.e., not squeezed) variances.

	\begin{subequations}
	We are able to characterize the input states using relations \eqref{eq:CovarRelA} and \eqref{eq:CovarRelB}.
	This yields the equivalent correlations for the bosonic operators from
	\label{eq:InitialCovars}
	\begin{eqnarray}
		\frac{v_{q,j}+v_{p,j}}{4}
		&&
		=\langle \Delta\hat{a}_j^\dag\Delta\hat{a}_j\rangle+\frac{1}{2}
		=\langle \Delta\hat{a}_j\Delta\hat{a}_j^{\dag}\rangle-\frac{1}{2},
		\\
		\frac{v_{q,j}-v_{p,j}}{4}
		&&
		=\langle \Delta\hat{a}_j^\dag\Delta\hat{a}_j^\dag\rangle
		=\langle \Delta\hat{a}_j\Delta\hat{a}_j\rangle.
	\end{eqnarray}
	\end{subequations}

\section{Algebraic results}\label{sec:GaussianCorr}

\subsection{Two-point correlator for Gaussian states}

	For our purposes, it is convenient to formulate the correlations in terms of central moments, $\hat{a}_j=\Delta\hat{a}_j+\alpha_{0,j}$, where $\alpha_{0,j}$ is the complex displacement [Eq. \eqref{eq:DisplRel}].
	This enables us to write the photon-number operators as
	\begin{equation}
		\hat{n}_j
		=\Delta\hat{a}_j^\dag\Delta\hat{a}_j
		+\Delta\hat{a}_j^\dag\alpha_{0,j}
		+\alpha_{0,j}^\ast\Delta\hat{a}_j
		+\alpha_{0,j}^\ast\alpha_{0,j}.
	\end{equation}
	Using this decomposition and $\langle\Delta\hat{a}_j\rangle=0$, the two-point correlators can be expanded as
	\begin{equation}
		\label{eq:CorrelationMatrixCentral}
	\begin{aligned}
		C_{j,k}
		=&\langle
			\Delta\hat a_j^\dag
			\Delta\hat a_j
			\Delta\hat a_k^\dag
			\Delta\hat a_k
		\rangle
		-\langle
			\Delta\hat a_j^\dag\Delta\hat a_j
		\rangle
		\langle
			\Delta\hat a_k^\dag\Delta\hat a_k
		\rangle
		\\
		&+\alpha_{0,j}\alpha_{0,k}^\ast
		\langle\Delta\hat a_j^\dag\Delta\hat a_k\rangle
		+\alpha_{0,j}^\ast\alpha_{0,k}
		\langle\Delta\hat a_j\Delta\hat a_k^\dag\rangle
		\\
		&+\alpha_{0,j}\alpha_{0,k}
		\langle\Delta\hat a_j^\dag\Delta\hat a_k^\dag\rangle
		+\alpha_{0,j}^\ast\alpha_{0,k}^\ast
		\langle\Delta\hat a_j\Delta\hat a_k\rangle
		\\
		&+\langle
			\big(\alpha_{0,j}^\ast\Delta\hat a_j+\alpha_{0,j}\Delta\hat a_j^\dag\big)
			\Delta\hat a_k^\dag\Delta\hat a_k
		\rangle
		\\
		&+\langle
			\Delta\hat a_j^\dag\Delta\hat a_j
			\big(\alpha_{0,k}^\ast\Delta\hat a_k+\alpha_{0,k}\Delta\hat a_k^\dag\big)
		\rangle.
	\end{aligned}
	\end{equation}

	For Gaussian states, the odd-order central moments vanish which means that the last two lines of Eq. \eqref{eq:CorrelationMatrixCentral} are zero.
	Moreover, the properties of Gaussian states imply that the first summand in Eq. \eqref{eq:CorrelationMatrixCentral} can be expressed in terms of second-order correlations (see the Appendix).
	Combining these considerations, we find that two-point correlators for Gaussian states read
	\begin{equation}
	\label{eq:GaussC}
	\begin{aligned}
		C_{j,k}
		{=}&\langle
			\Delta\hat a_j^\dag
			\Delta\hat a_k^\dag
		\rangle
		\langle
			\Delta\hat a_j
			\Delta\hat a_k
		\rangle
		+\langle
			\Delta\hat a_j^\dag
			\Delta\hat a_k
		\rangle
		\langle
			\Delta\hat a_j
			\Delta\hat a_k^\dag
		\rangle
		\\
		&+\alpha_{0,j}\alpha_{0,k}^\ast
		\langle\Delta\hat a_j^\dag\Delta\hat a_k\rangle
		+\alpha_{0,j}^\ast\alpha_{0,k}
		\langle\Delta\hat a_j\Delta\hat a_k^\dag\rangle
		\\
		&+\alpha_{0,j}\alpha_{0,k}
		\langle\Delta\hat a_j^\dag\Delta\hat a_k^\dag\rangle
		+\alpha_{0,j}^\ast\alpha_{0,k}^\ast
		\langle\Delta\hat a_j\Delta\hat a_k\rangle.
	\end{aligned}
	\end{equation}
	It is worth noting that for a vanishing displacement, only the first line in Eq. \eqref{eq:GaussC} contributes.

\subsection{Propagation in the interferometer}

	In the following, let us describe the propagation of two-point correlators in the optical network.
	The relations \eqref{eq:InitialCovars} can describe the bosonic ladder-operator correlations for the initial state, and Eq. \eqref{eq:GaussC} gives their relation to the photon-number correlators.
	Thus, for the family of Gaussian initial states under consideration, we get correlators of the form $C_{j,k}^\mathrm{(in)}=0$ for $j\neq k$ and
	\begin{equation}
		C_{j,j}^\mathrm{(in)}
		=\frac{v_{q,j}^2+v_{p,j}^2-2+2\xi_{0,j}^2v_{q,j}+2\xi_{0,j+M}^2v_{p,j}}{8}.
	\end{equation}

	Further, we recall that the propagation in the network yields $\hat{\boldsymbol a}\mapsto\boldsymbol U\hat{\boldsymbol a}$.
	Then the definition of the two-point correlator results in the following evolution of the input correlators:
	\begin{equation}
	\begin{aligned}
		C_{j,k}^\mathrm{(out)}
		=& \sum_{r,s,t,u=1}^M
		U_{j,r}^\ast U_{j,s} U_{k,t}^\ast U_{k,u}
		\\
		&{\times}\Big(
			\langle \hat a_{r}^\dag\hat a_{s}\hat a_t^\dag\hat a_{u}\rangle^\mathrm{(in)}
			-\langle \hat a_{r}^\dag\hat a_{s}\rangle^\mathrm{(in)}\langle \hat a_t^\dag\hat a_{u}\rangle^\mathrm{(in)}
		\Big).
	\end{aligned}
	\end{equation}
	Here the superscripts ``(in)'' and ``(out)'' are introduced to clearly differentiate between the input and output modes, respectively.
	As demonstrated above, we can again write the input correlations in terms of central moments and use the properties of central moments of Gaussian states (cf. the Appendix).
	Consequently, we find
	\begin{eqnarray}
		&&\langle \hat a_{r}^\dag\hat a_{s}\hat a_t^\dag\hat a_{u}\rangle^\mathrm{(in)}
		-\langle \hat a_{r}^\dag\hat a_{s}\rangle^\mathrm{(in)}\langle \hat a_t^\dag\hat a_{u}\rangle^\mathrm{(in)}
		\\\nonumber
		&{=}&\delta_{s,t}\delta_{r,u}
		\langle \Delta\hat a_s\Delta\hat a_s^\dag\rangle^\mathrm{(in)}
		\langle \Delta\hat a_r^\dag\Delta\hat a_r\rangle^\mathrm{(in)}
		\\\nonumber
		&&{+}\delta_{r,t}\delta_{s,u}
		\langle \Delta\hat a_r^\dag\Delta\hat a_r^\dag\rangle^\mathrm{(in)}
		\langle \Delta\hat a_s\Delta\hat a_s\rangle^\mathrm{(in)}
		\\\nonumber
		&&{+}\delta_{s,t}\alpha_{0,r}^\ast\alpha_{0,u}\langle \Delta\hat a_s\Delta\hat a_s^\dag\rangle^\mathrm{(in)}
		{+}\delta_{r,u}\alpha_{0,s}\alpha_{0,t}^\ast\langle \Delta\hat a_r^\dag\Delta\hat a_r\rangle^\mathrm{(in)}
		\\\nonumber
		&&{+}\delta_{s,u}\alpha_{0,t}^\ast\alpha_{0,r}^\ast\langle \Delta\hat a_s\Delta\hat a_s\rangle^\mathrm{(in)}
		{+}\delta_{r,t}\alpha_{0,s}\alpha_{0,u}\langle \Delta\hat a_r^\dag\Delta\hat a_r^\dag\rangle^\mathrm{(in)},
	\end{eqnarray}
	also using that there are no cross correlations in the input state, $\langle \Delta \hat a_{x}^\dag\Delta\hat a_{y}\rangle^\mathrm{(in)}=0=\langle\Delta \hat a_x\Delta\hat a_{y}\rangle^\mathrm{(in)}$ for $x\neq y$.
	Inserting this into the previous relation, we obtain
	\begin{equation}
	\label{eq:InOutGen}
	\begin{aligned}
		C_{j,k}^\mathrm{(out)}
		=&
		\left(S_{j,k}[U]^\ast+\frac{1}{2}\delta_{j,k}\right)
		\left(S_{j,k}[U]-\frac{1}{2}\delta_{j,k}\right)
		\\
		&+D_{j,k}[U]D_{j,k}[U]^\ast
		\\
		&+\alpha_{U,j}^\ast\alpha_{U,k} S_{j,k}[U]^\ast
		+\alpha_{U,j}\alpha_{U,k}^\ast S_{j,k}[U]
		\\
		&+\alpha_{U,j}^\ast\alpha_{U,k}^\ast D_{j,k}[U]
		+\alpha_{U,j}\alpha_{U,k} D_{j,k}[U]^\ast,
	\end{aligned}
	\end{equation}
	\begin{subequations}
	with the abbreviations $\alpha_{U,l}=\sum_{w=1}^M U_{l,w}\alpha_{0,w}$, such that $\alpha_{U,l}$ is the coherent component of the output mode $l$, and
	\begin{eqnarray}
		S_{j,k}[U]
		&=&\sum_{w=1}^M U_{j,w}^\ast U_{k,w}\frac{
			v_{q,w}+v_{p,w}
		}{4},
		\\
		D_{j,k}[U]
		&=&\sum_{w=1}^M U_{j,w}U_{k,w}\frac{
			v_{q,w}-v_{p,w}
		}{4}.
	\end{eqnarray}
	\end{subequations}

	The finding in Eq. \eqref{eq:InOutGen} presents the most general input-output relation of two-point, photon-number correlators for the scenario of Gaussian boson sampling with independent inputs.
	From now on, we exclusively focus on the output correlations.
	Therefore, we skip the superscript and $C_{j,k}$ exclusively refers to the output correlations in all following considerations.
	Also note in this context that the input is uniquely defined by the input variances $v_{q,j}$ and $v_{p,j}$ as well as the displacement vector $\boldsymbol \xi_0$.

\subsection{Discussion}

	For a known unitary and a well-characterized input state, Eq. \eqref{eq:InOutGen} can be directly evaluated.
	For example, when coherent states are injected (i.e., $v_p = v_q = 1$), we immediately obtain $C_{j,k} = 0$.
	The result in Eq. \eqref{eq:InOutGen} is generally useful for simulating experiments in which a mean field is present.
	Yet, additional terms, associated with the displacement in phase space, considerably increase the complexity of random matrix calculations.
	Moreover, displacements (e.g., by mixing states on a beam splitter with a coherent state) are hard to generate in the experimental setting of interest.
	Also note that displacement is a classical operation, which makes it an unlikely resource for a quantum advantage.
	Therefore, in the remainder of this article, we focus on nondisplaced input states and set $\alpha_{0,w}=0$ for all input modes $w = 1, \dots, M$.
	It is then practical to recast Eq. \eqref{eq:InOutGen} in a form that explicitly captures the structure of the correlations in terms of the components of the unitary circuit,
	\begin{eqnarray}
		\label{eq:InOutUsed}
		&& C_{j,k}
		\\\nonumber
		&=& \sum_{w,w'=1}^M  \frac{(v_{q,w}{+}v_{p,w})(v_{q,w'}{+}v_{p,w'})}{16} U_{j,w} U_{k,w'}U_{j,w'}^\ast U_{k,w}^\ast 
		\\\nonumber
		&& {+}\sum_{w,w'=1}^M  \frac{(v_{q,w}{-}v_{p,w})(v_{q,w'}{-}v_{p,w'})}{16} U_{j,w} U_{k,w}U_{j,w'}^\ast U_{k,w'}^\ast
		\\\nonumber
		&& -\frac{1}{4}\delta_{j,k}.
	\end{eqnarray}

	This result not only provides an interesting tool for benchmarking experiments; it also offers a direct comparison to many-boson interference using Fock states as inputs \cite{WKUMTRB16,MTMKB11}.
	In such arrangements, the correlators $C_{j,k}$ are associated solely with two-particle interference processes, arising from terms proportional to $U_{j,w} U_{k,w'}U_{j,w'}^\ast U_{k,w}^\ast$.
	These terms also appear in Eq. \eqref{eq:InOutUsed} and, thus, can be considered hallmarks of similar interference processes appearing in the present Gaussian setting.
	However, there are also considerable differences between the Fock state correlators \cite{WKUMTRB16,MTMKB11} and the Gaussian scenario in Eq. \eqref{eq:InOutUsed}.
	For instance, the terms proportional to $U_{j,w} U_{k,w}U_{j,w}^\ast U_{k,w}^\ast$ (for $w'=w$) are added in the Gaussian case, whereas they are subtracted in the Fock state case.

	Even more profound is the appearance of a completely new class of terms proportional to $U_{j,w} U_{k,w}U_{j,w'}^\ast U_{k,w'}^\ast$, which are absent in scenarios with fixed particle numbers.
	As their contribution is weighted with the difference of the variances, they reflect the nonrotational invariance of the initial Gaussian states when compared to Fock states (and mixtures thereof) in phase space.
	The appearance of this new class of terms may indicate the presence of a new type of interference phenomenon that can manifest itself in Gaussian boson sampling.
	In particular, this is an indication that Gaussian interference experiments may show new physics beyond the many-particle interference processes for boson sampling with Fock states.

	Thus, with the aim of quantifying the impact of phase-dependent input states, it is sensible to introduce the operators
	\begin{equation}
		\label{eq:EccentricityDef}
		\hat \epsilon_j=\frac{\hat q_j^2-\hat p_j^2}{4}
	\end{equation}
	for $j=1,\ldots,M$, which have in our scenario the expectation values $\langle\hat\epsilon_j\rangle=(v_{q,j}-v_{p,j})/4$.
	This quantity is the difference of the two quadratures and characterizes the eccentricity of the uncertainty ellipse in phase space.
	Also, note that $\hat \epsilon_j$ complements the definition of the photon-number operator, $\hat n_j+1/2=(\hat q_j^2+\hat p_j^2)/4$.

\subsection{Randomization over unitaries}

	It is possible to use random matrix theory to obtain analytical expressions for $\mathbb{E}_{U}\left(C\right)$, $\mathbb{E}_{U}\left(C^{2}\right)$, and $\mathbb{E}_{U}\left(C^{3}\right)$ to evaluate the quantities in \eqref{eq:RandomMatrixStatProp}, defining $\mathsf{NM}$, $\mathsf{CV}$, and $\mathsf{Sk}$.
	The randomization yields the same result when swapping output modes which corresponds to a unitary transformation, mapping the set of unitaries onto itself and, thus, does not affect the Haar measure.
	This justifies the notation $\mathbb E_U(C_{j,k}^x)=\mathbb E_U(C^x)$ for any integer $x$ and $j\neq k$.

	We focus on a scenario with $N$ occupied modes, implying $M-N$ vacuum inputs ($1\leq N\leq M$).
	As permutations of input modes are unitary operations, we further say that the first $N$ modes are the occupied ones.
	Further on, like in the case of boson sampling with single photons, we assume that the states in the occupied modes are all identical.
	Thus, we have the input quadrature variances ($s=p,q$)
	\begin{equation}
	\label{eq:veff}
		v_{s,j}=\left\lbrace\begin{array}{lll}
			v_s & \text{ for } & j=1,\ldots,N,
			\\
			1 & \text{ for } & j=N+1,\ldots,M.
		\end{array}\right.
	\end{equation}
	Furthermore, the average photon number in the occupied input modes is given by
	\begin{equation}
	\label{eq:neff}
		\langle \hat{n}_j\rangle=\langle \hat{n}\rangle = \frac{v_q + v_p - 2}{4}.
	\end{equation}
	In addition, the eccentricity [Eq. \eqref{eq:EccentricityDef}] reads
	\begin{equation}
		\langle \hat\epsilon_j\rangle=\langle\hat\epsilon\rangle=\frac{v_q-v_p}{4}
	\end{equation}
	for the $N$ occupied input modes.
	With these considerations, we obtain
	\begin{equation}
	\label{eq:InOutSimpler}
	\begin{aligned}
		C_{j,k}
		=&  \langle \hat{n}\rangle^2 \sum_{w,w'=1}^N  U_{j,w} U_{k,w'}U_{j,w'}^\ast U_{k,w}^\ast 
		\\
		& +\langle\hat\epsilon\rangle^2\sum_{w,w'=1}^N U_{j,w} U_{k,w}U_{j,w'}^\ast U_{k,w'}^\ast.
	\end{aligned}
	\end{equation}

	To evaluate the random matrix average $\mathbb{E}_{U}\left(C\right)$, we use the linearity of the expectation value such that
	\begin{equation}
	\label{eq:InOutAV}
	\begin{split}
		\mathbb{E}_{U}\left(C\right)
		=&  \langle \hat{n}\rangle^2 \sum_{w,w'=1}^N  \mathbb{E}_{U}\left(U_{j,w} U_{k,w'}U_{j,w'}^\ast U_{k,w}^\ast \right)
		\\
		& +\langle\hat\epsilon\rangle^2\sum_{w,w'=1}^N \mathbb{E}_{U}\left(U_{j,w} U_{k,w}U_{j,w'}^\ast U_{k,w'}^\ast\right).
	\end{split}
	\end{equation}
	The averages can then be obtained through the following identity for $M\times M$ random unitary matrices $\boldsymbol U$ \cite{C78,W78,S80}:
	\begin{eqnarray}
	\label{eq:UnitaryAverage}
	         &&\mathbb{E}_U(U_{a_1,b_1}\dots U_{a_n,b_n}U^*_{\alpha_1,\beta_1}\dots U^*_{\alpha_n,\beta_n}) 
	         \\\notag
	         & = & \sum_{\sigma,\pi \in S_n}\mathcal V_M(\sigma^{-1}\pi)\prod^n_{k=1}\delta(a_k-\alpha_{\sigma(k)})\delta(b_k-\beta_{\pi(k)}),
	\end{eqnarray}
	where $S_n$ denotes the permutation group for $n$ elements and $\mathcal V$ are class coefficients (also known as Weingarten functions), typically determined recursively.
	Because only low-order terms are considered here, the necessary values for the class coefficients can be taken from the literature \cite{BB96}.
	For higher order moments, it is often convenient to resort to alternatives, such as semiclassical methods \cite{BK13}, or use a direct, yet sophisticated approach based on the Schur-Weyl duality \cite{CS06}.

	Furthermore, for the evaluation of the higher moments $\mathbb{E}_{U}\left(C^2\right)$ and $\mathbb{E}_{U}\left(C^3\right)$, it suffices to straightforwardly evaluate $C_{jk}^2$ and $C_{jk}^3$ and apply the same techniques.
	These computations will rapidly become more intricate because of the appearance of cross terms, which introduce new types of nonvanishing terms when applying Eq. \eqref{eq:UnitaryAverage}.

\begin{widetext}
	To implement relation \eqref{eq:UnitaryAverage} and do the bookkeeping of indices in the calculation of $\mathbb{E}_{U}\left(C\right)$, $\mathbb{E}_{U}\left(C^2\right)$, and $\mathbb{E}_{U}\left(C^3\right)$, we resort to methods that are analogous to those detailed in Appendix B of Ref. \cite{W16}.
	\begin{subequations}
	By summing all the nonzero contributions upon evaluation of Eq. \eqref{eq:UnitaryAverage}, we obtain as the final result
	\label{eq:CMoments}
	\begin{eqnarray}
		\mathbb{E}_{U}(C)
		&=&\frac{N(M-N) }{(M-1)M(M+1)}\langle \hat{n}\rangle^2
		+\frac{N}{M(M+1)}\langle\hat\epsilon\rangle^2,
	\\\nonumber
		\mathbb{E}_{U}(C^2)
		&=&\frac{2 N (N+1) (M-N+1) (M-N)}{(M-1) M^2 (M+1) (M+2) (M+3)}\langle \hat{n}\rangle^4
		+\frac{2 N (M-N) (MN+3M-N+1)}{(M-1) M^2 (M+1) (M+2) (M+3)}\langle \hat{n}\rangle^2 \langle\hat\epsilon\rangle^2
		\\
		&&+\frac{2 N  ( M^2 N+M^2+NM-3M+ 2N-2)}{(M-1) M^2 (M+1) (M+2) (M+3)}\langle\hat\epsilon\rangle^4,
	\\
		\nonumber
		\mathbb{E}_{U}(C^3)
		&=&\frac{6 (N+1) N (N+2) (M-N+2) (M-N+1) (M-N)}{(M-1) M^2 (M+1)^2 (M+2) (M+3) (M+4) (M+5)} \langle \hat{n}\rangle^6
		\\\nonumber
		&&+\frac{6 N (N+2) (M-N) (M-N+1) (MN+5M-N+7)}{(M-1) M^2 (M+1)^2 (M+2) (M+3) (M+4) (M+5)} \langle \hat{n}\rangle^4\langle\hat\epsilon\rangle^2
		\\\nonumber
		&&+\frac{6 N (N+2) (M-N) ( M^2N+MN+5M^2+5M+4N-4)}{(M-1) M^2 (M+1)^2 (M+2) (M+3) (M+4) (M+5)} \langle \hat{n}\rangle^2\langle\hat\epsilon\rangle^4
		\\
		&&+\frac{6 N (N+2) (M^2N+5M N+M^2-7M+12N-12)}{(M-1) M^2 (M+1) (M+2) (M+3) (M+4) (M+5)} \langle\hat\epsilon\rangle^6.
	\end{eqnarray}
	\end{subequations}
	These expressions can then be inserted into Eq. \eqref{eq:RandomMatrixStatProp} to straightforwardly obtain analytical expressions for $\mathsf{NM}$, $\mathsf{CV}$, and $\mathsf{Sk}$.
	Also, in the next section, we use numerical methods to investigate how these analytical predictions compare to simulated outcomes for a Gaussian boson sampling experiment.
\end{widetext}

\section{Simulation results}\label{sec:Simulations}

	In the following, we simulate an experimental setup with reconfigurable linear optics and photon-number-resolved detection and investigate the impact of the properties of the input states on the benchmarking scheme.
	Specifically, we study thermal and squeezed states (cf. the discussion at the end of Sec. \ref{sec:Gaussian}) as two paradigmatic examples of relevance for experimental implementations.
	In addition, for all simulations and without loss of generality, we set $j=1$ and $k=2$, meaning that we are working with $C_{1,2}$ the whole time.

\subsection{Simulation methods} \label{SimMeth}

	Two different methods can be used to simulate the values of $C_{1,2}$ for different Haar-random unitaries.
	The first one is closest to what would be done in a laboratory.
	We first use Eqs. \eqref{eq:DispVec} and \eqref{eq:CovMat} to get the covariance matrix $\boldsymbol{V}$ and displacement vector $\boldsymbol{\xi}_0$ for the state under consideration.
	We then use Eqs. \eqref{eq:VTrans} and \eqref{eq:UonV} for the unitary evolution.
	The following step is tracing over all but modes 1 and 2; that is, we only consider the $4\times 4$ matrix and four-component vector
	\begin{equation}
		\tilde{\boldsymbol V}=\left(\begin{array}{cccc}
			V_{1,1} & V_{1,2} & V_{1,M+1} & V_{1,M+2}
		\\
			V_{2,1} & V_{2,2} & V_{2,M+1} & V_{2,M+2}
		\\
			V_{M+1,1} & V_{M+1,2} & V_{M+1,M+1} & V_{M+1,M+2}
		\\
			V_{M+2,1} & V_{M+2,2} & V_{M+2,M+1} & V_{M+2,M+2}
		\end{array}\right)
	\end{equation}
	and $\tilde{\boldsymbol\xi}_0=\left(
			\xi_{0,1},
			\xi_{0,2},
			\xi_{0,M+1},
			\xi_{0,M+2}
		\right)^\mathrm{T}$ of the output state.
	With those, we then compute the photon-number distribution using the procedures in Refs. \cite{DMM94} and \cite{KB01}.
	This gives an array of values for the probabilities $P\left(n_{1},n_{2}\right)$ of detecting $\left(n_{1},n_{2}\right)$ photons; then, $C_{1,2}$ is directly calculated.
	This is a straightforward, yet a highly computationally inefficient approach as it corresponds to simulating Gaussian boson sampling, a problem considered to be computationally difficult \cite{HKSBSJ17,KHSBSJ18}.
	As Gaussian states do not have a finite photon-number distribution---though the probabilities of detecting higher photon numbers get increasingly smaller---a maximum photon-number resolution $n_{\max}$ can be defined.
	This has implications for $C_{1,2}$ which are discussed in detail in Sec. \ref{sec:DetectorPNR}.

	The other method of simulating $C_{1,2}$ is to use our results from Sec. \ref{sec:GaussianCorr} directly.
	This approach is much quicker as it avoids the intermediate calculation of photon-number distributions \cite{DMM94,KB01}.
	From the set of randomly generated $C_{1,2}$ values, $\mathsf{NM}$, $\mathsf{CV}$ and $\mathsf{Sk}$ are obtained.
	These values can be compared to the exact values in Eqs. \eqref{eq:RandomMatrixStatProp} and \eqref{eq:CMoments} for the same systems to get an idea of how many Haar-random unitary evolutions one requires to determine good estimates for $\mathsf{NM}$, $\mathsf{CV}$ and $\mathsf{Sk}$ in simulations and future experiments.

\subsection{Squeezed and thermal state comparison} \label{SqueezeThermalComp}

	Pure squeezed states form a class of Gaussian states that cannot be modeled with classical light.
	They have been produced in experiments for decades and thus serve as a good starting point to develop an intuition for our benchmarking scheme.
	Furthermore, squeezed-vacuum inputs are the archetypal scenario for Gaussian boson sampling \cite{HKSBSJ17,KHSBSJ18}.
	In contrast, thermal light behaves in a highly classical way, rendering it an ideal example to contrast against the squeezed vacuum.
	It is worth emphasizing that Gaussian boson sampling with thermal input states can be simulated in an efficient way.

\subsubsection{Small and large systems}

	First, we separately consider squeezed and thermal states as inputs for boson sampling to gather insights into their characteristic features.
	This is done using a small system of $M=8$ and $N=2$, typical for what is currently achievable in a laboratory, as well as in a large system of $M=120$ and $N=10$ to compare with the results from Ref. \cite{WKUMTRB16}.
	These numbers were selected due to the technical requirement in boson sampling of having many more modes available than nonvacuum input states, i.e., $M\gtrsim N^{\nu}$ for $\nu=2$.
	As it was shown that boson sampling cannot be hard for $\nu<2$ \cite{AA13}, we specifically focus on this borderline case.

	Complementary to the definition in Sec. \ref{sec:Gaussian}, based on the state's covariance matrix, it is helpful to expand the squeezed and thermal states in the photon-number basis for further insight.
	A single-mode squeezed state $|S\rangle $ (without displacement) is described through the squeezing operator acting on the vacuum state, i.e., $|S\rangle = \exp\left[r\left(e^{i\phi}\hat a^{\dagger 2}-e^{-i\phi}\hat{a}^{2}\right)/2\right] |0\rangle$, where $r$ is the squeezing parameter and $\phi$ defines the antisqueezing axis in phase space.
	Thus, the squeezed state exhibits the photon-number basis expansion
	\begin{equation}
 	\label{eq:SqueezeFock}
		|S\rangle
		=\frac{1}{\sqrt{\cosh (r)}}\sum_{n=0}^\infty
		\left(e^{i\phi}\tanh (r)\right)^{n}\frac{\sqrt{\left(2n\right)!}}{n!2^{n}}|2n\rangle.
	\end{equation}
	In this form, the squeezed state is a coherent superposition of photon-number states.
	When this is generalized to a squeezed state input in $M$ modes, the covariance matrix is given by
	\begin{equation}
	\label{eq:SqueezeCov}
		\boldsymbol{V}=\mathrm{diag}\left(
				e^{2r_{1}}, \ldots, e^{2r_{M}}, e^{-2r_{1}}, \ldots, e^{-2r_{M}}
		\right)
	\end{equation}
	where $r_j$ is the squeezing parameter for mode $j$ (note that $r_j=0$ corresponds to vacuum in mode $j$).
	Likewise, we have $v_{q,j}=e^{2r_{j}}$ and $v_{p,j}=e^{-2r_{j}}$ in Eq. \eqref{eq:InitialCovars}.
	Thus, we get $\langle\hat n_j\rangle=\sinh^{2}(r_j)$, and the eccentricity is quantified as $\langle\hat \epsilon_j\rangle=\sinh(2r_j)/2$ [Eq. \eqref{eq:EccentricityDef}].
	As a local diagonalization can be performed, and we can choose squeezing along the $p$ quadrature axis and antisqueezing along the $q$ quadrature axis, we set $\phi=0$ in Eq. \eqref{eq:SqueezeFock}.

	By contrast, a thermal state $\hat{\rho}_{T}$ is a classical (i.e., incoherent) mixture of photons,
	\begin{equation}
	\label{eq:ThermalFock}
		\hat{\rho}_{T}
		=\frac{1}{\bar n +1}\sum_{n=0}^\infty
		\left(\frac{\bar n }{\bar n +1}\right)^{n}|n\rangle \langle n|,
	\end{equation}
	where $\bar n=\langle\hat n\rangle$ is the mean thermal photon number.
	For a thermal state, the off-diagonal density matrix elements in the Fock basis are always zero as thermal states are rotationally invariant, also implying $\langle\hat \epsilon\rangle=0$.
	Again, when generalizing this to $M$ modes, we obtain
	\begin{equation}
		\label{eq:ThermalCov}
		\boldsymbol{V}=\mathrm{diag}\left(
			2\bar n_1{+}1,\ldots,2\bar n_M{+}1,
			2\bar n_1{+}1,\ldots,2\bar n_M{+}1
		\right),
	\end{equation}
	with $\bar n_j$ denoting the mean photon number of the $j$th mode.
	Let us stress that $\bar n_j=0$ corresponds to a vacuum input state in mode $j$.

\begin{figure}[tb]
	\includegraphics[width=\columnwidth]{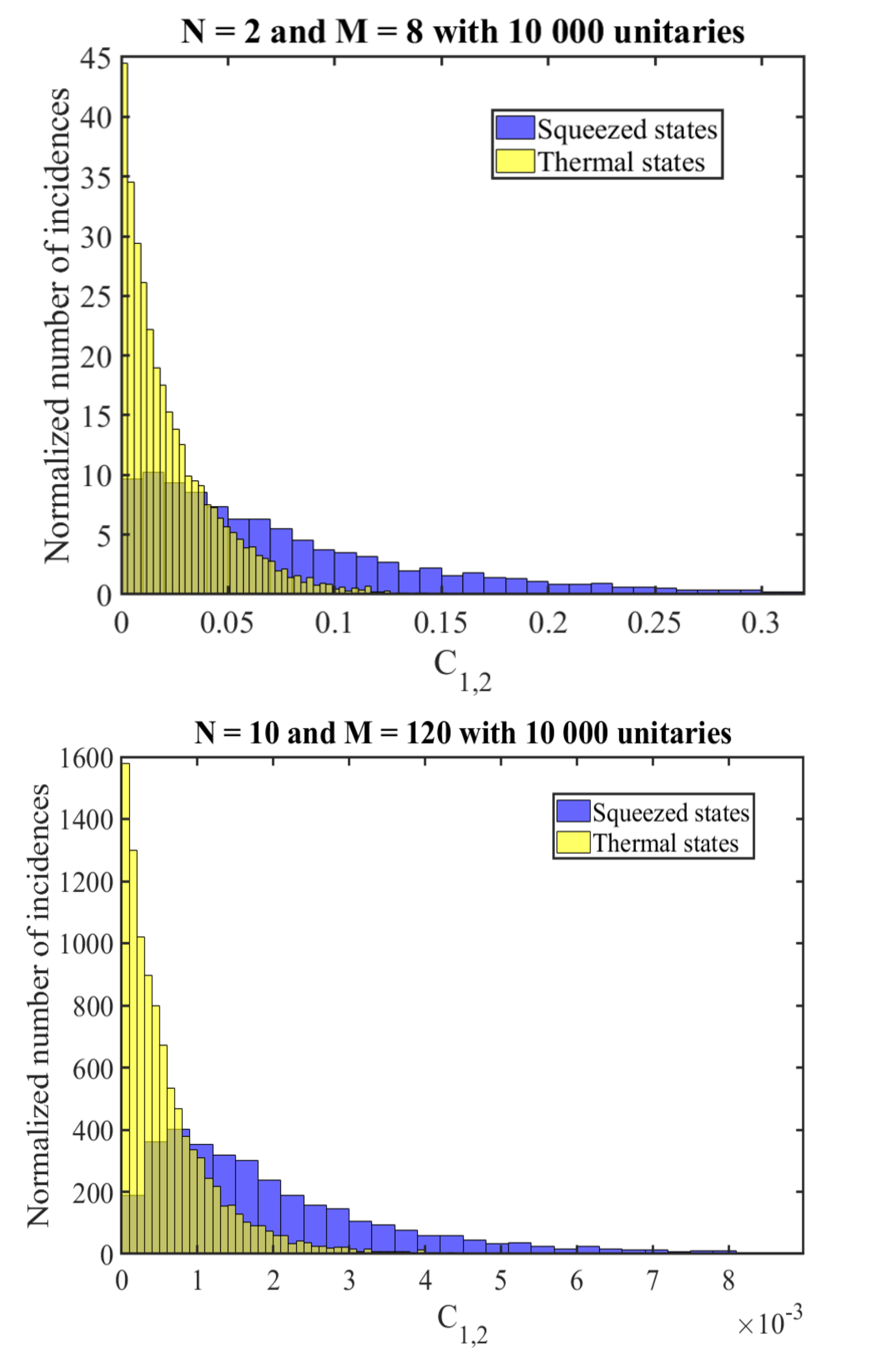} 
	\caption{(Color online)
		Comparison of histograms of $C_{1,2}$ for squeezed states [$r=\mathrm{ln}(1+\sqrt{2})$ ($\langle \hat{n}\rangle=1$) and $\phi=0$; cf. Eq. \eqref{eq:SqueezeFock}] and thermal states [$\bar n=1$; cf. Eq. \eqref{eq:ThermalFock}].
		For both plots, a sample of $10\,000$ different Haar-random unitaries was generated.
		The top plot shows the histogram for two occupied modes out of eight available modes, $N=2$ and $M=8$, respectively.
		In the bottom plot, we have $N=10$ and $M=120$.
	}\label{fig:SmallLargeC12Hists}
\end{figure}

	In our simulations, we consider the scenario of $M$ identical input modes [cf. Eq. \eqref{eq:veff}].
	This leads to typical histograms for squeezed and thermal states for small and large systems as shown in Fig. \ref{fig:SmallLargeC12Hists}, top and bottom plots, respectively.
	In both cases, we have the same mean photon number per mode for the input states, $\langle\hat n\rangle=1$.
	The eccentricity for the thermal state is zero, whereas we have $\langle\hat\epsilon\rangle=\sqrt{2}$ for the squeezed state.
	Because of the latter phase dependence, we have an additional contribution to $C_{1,2}$ [cf. Eq. \eqref{eq:InOutAV}], resulting in a distinctively broader distribution for squeezed states compared to thermal states with the same number of photons.

\begin{figure*}
	\includegraphics[width=1\textwidth]{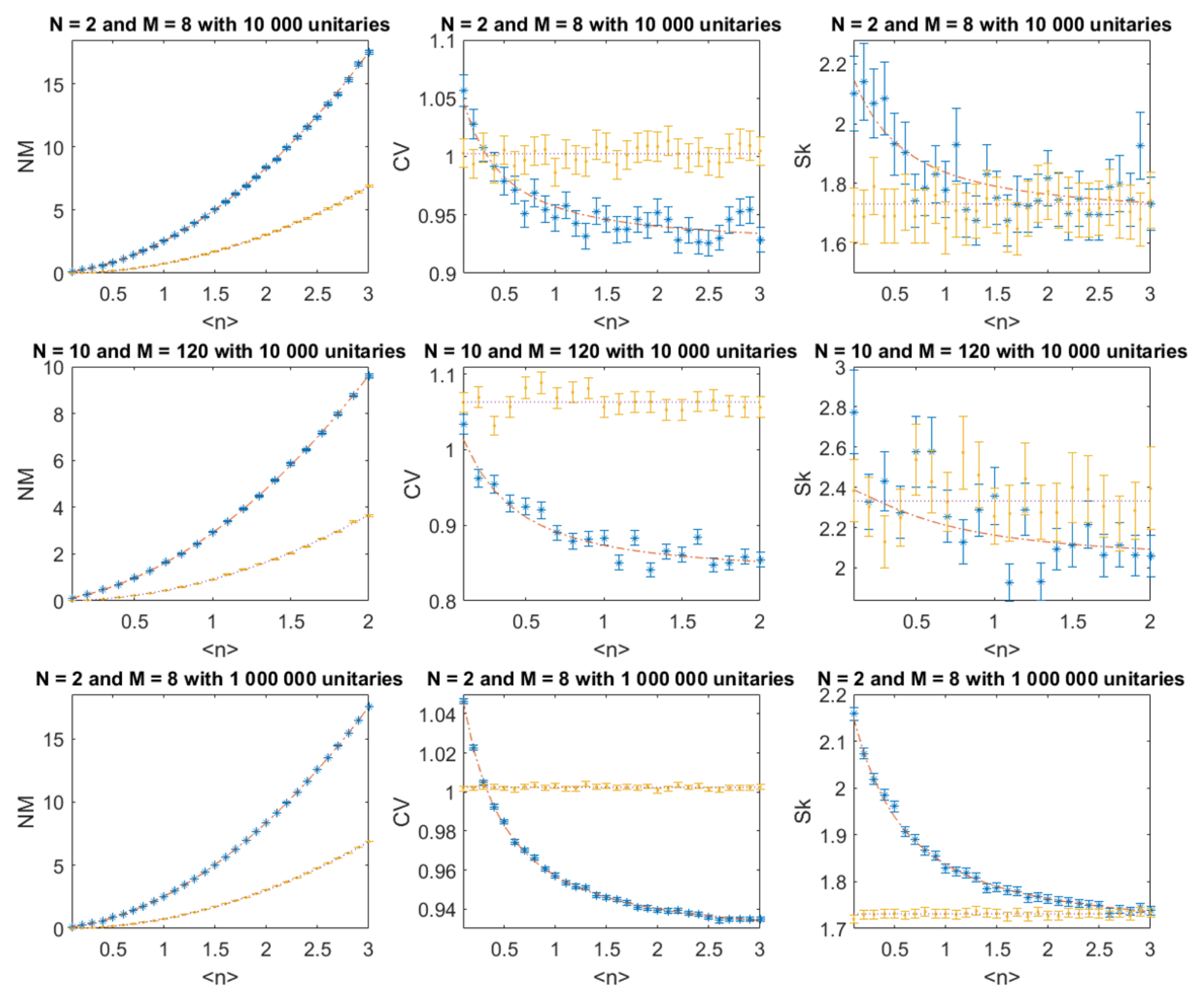}
	\caption{(Color online)
		Parameters $\mathsf{MN}$, $\mathsf{CV}$, and $\mathsf{Sk}$ (columns from left to right) for squeezed and thermal states.
		The analytical expressions (orange dot-dashed lines for squeezed states and purple dotted lines for thermal states) and the values obtained from the simulated data with $1\sigma$ error bars (blue stars for squeezed states and yellow dots for thermal states) are plotted.
		The top row shows the results for a small system, $N=2$ and $M=8$, and $10\,000$  Haar-random unitaries are generated to sample $C_{1,2}$.
		The middle row shows the results for a large system, $N=10$ and $M=120$, and a Haar-random sample of $10\,000$ values for $C_{1,2}$ from $10\,000$.
		The bottom row shows the results for a small system, $N=2$ and $M=8$, however, for an increased sample size of $1\,000\,000$ values for $C_{1,2}$ compared to the first row.
	}\label{fig:NMCVSk}
\end{figure*}

	In Fig. \ref{fig:NMCVSk}, we compare the variation of $\mathsf{NM}$, $\mathsf{CV}$ and $\mathsf{Sk}$, as defined in Eqs. \eqref{eq:RandomMatrixStatProp}, for both types of state and for large and small systems.
	To do so, we average over the simulation values of $C_{1,2}$, $C_{1,2}^{2}$, and $C_{1,2}^{3}$, and compare the results to the values of $\mathsf{NM}$, $\mathsf{CV}$ and $\mathsf{Sk}$ that are obtained via the relations in Eq. \eqref{eq:CMoments}.
	On the one hand, these results allow us to probe and, thereby, distinguish the features of squeezed and thermal states.
	For thermal states, $\mathsf{CV}$ and $\mathsf{Sk}$ are constant with varying average photon number which can be understood from Eq. \eqref{eq:CMoments} for $\langle\hat\epsilon\rangle=0$.
	Therefore, the $\langle \hat{n}\rangle $ terms cancel out in the final expression for $\mathsf{CV}$ and $\mathsf{Sk}$.
	For squeezed states, we do observe an effect of altering $\langle \hat{n}\rangle$, which can be used as a method to distinguish both types of states.

	Moreover, through all of Fig. \ref{fig:NMCVSk}, we gauge the number of iterations that are required to let the statistics of the simulated data converge to the analytical predictions as marked by the standard error.
	Because of the increased standard error, it is obvious that the uncertainties are larger in Fig. \ref{fig:NMCVSk} (top) compared to the corresponding plot in Fig. \ref{fig:NMCVSk} (bottom), describing $10\,000$ iterations versus $1\,000\,000$, respectively.
	Furthermore, by comparing Fig. \ref{fig:NMCVSk} (top) and Fig. \ref{fig:NMCVSk} (middle), we can observe that the system size does not affect the relative uncertainties for the corresponding moment for the same number of iterations.
	This demonstrates that we need to be conscious of the number of iterations performed depending on which moment we wish to consider, though with a tunable photonic circuit, it would be experimentally possible to generate sufficiently many random unitaries to reach the necessary statistical error on the measured data.

	From the different results in Fig. \ref{fig:NMCVSk}, it seem appropriate to use $\mathsf{NM}$ to distinguish between squeezed and thermal states at large $\langle \hat{n}\rangle $ and $\mathsf{Sk}$ to tell them apart at small $\langle \hat{n}\rangle$.
	However, due to the relation $\sigma_{\mathsf{NM}}<\sigma_{\mathsf{CV}}<\sigma_{\mathsf{Sk}}$ (where $\sigma_{x}$ is the standard error in $x\in\{\mathsf{NM,CV,Sk}\}$), it becomes apparent that it is most efficient to use $\mathsf{NM}$ to discriminate between squeezed and thermal input states.
	Also there will always be implementation-dependent sources of error in addition to the statistical uncertainties; therefore further error analysis directly on experimental data would be required \cite{Getal18}.
	This is discussed in more detail in Sec. \ref{sec:DiscStatSig}.

\subsubsection{Constant dilution}

	So far, we studied the impact of the type of state and the sample size on the implementation of boson sampling protocols with Gaussian states.
	We now investigate the influence of the distribution of a fixed amount of energy (i.e., total photon number) into a varying number of occupied modes, referred to as constant dilution.
	The motivation to study such a problem comes from Ref. \cite{KHSBSJ18}, where the impact of multiphoton events in the same mode is considered.
	Specifically, the question is addressed whether it is more favorable to increase the squeezing and use a few occupied inputs or have less squeezing distributed over a larger number of modes.

\begin{figure}[tb]
	\includegraphics[width=1\columnwidth]{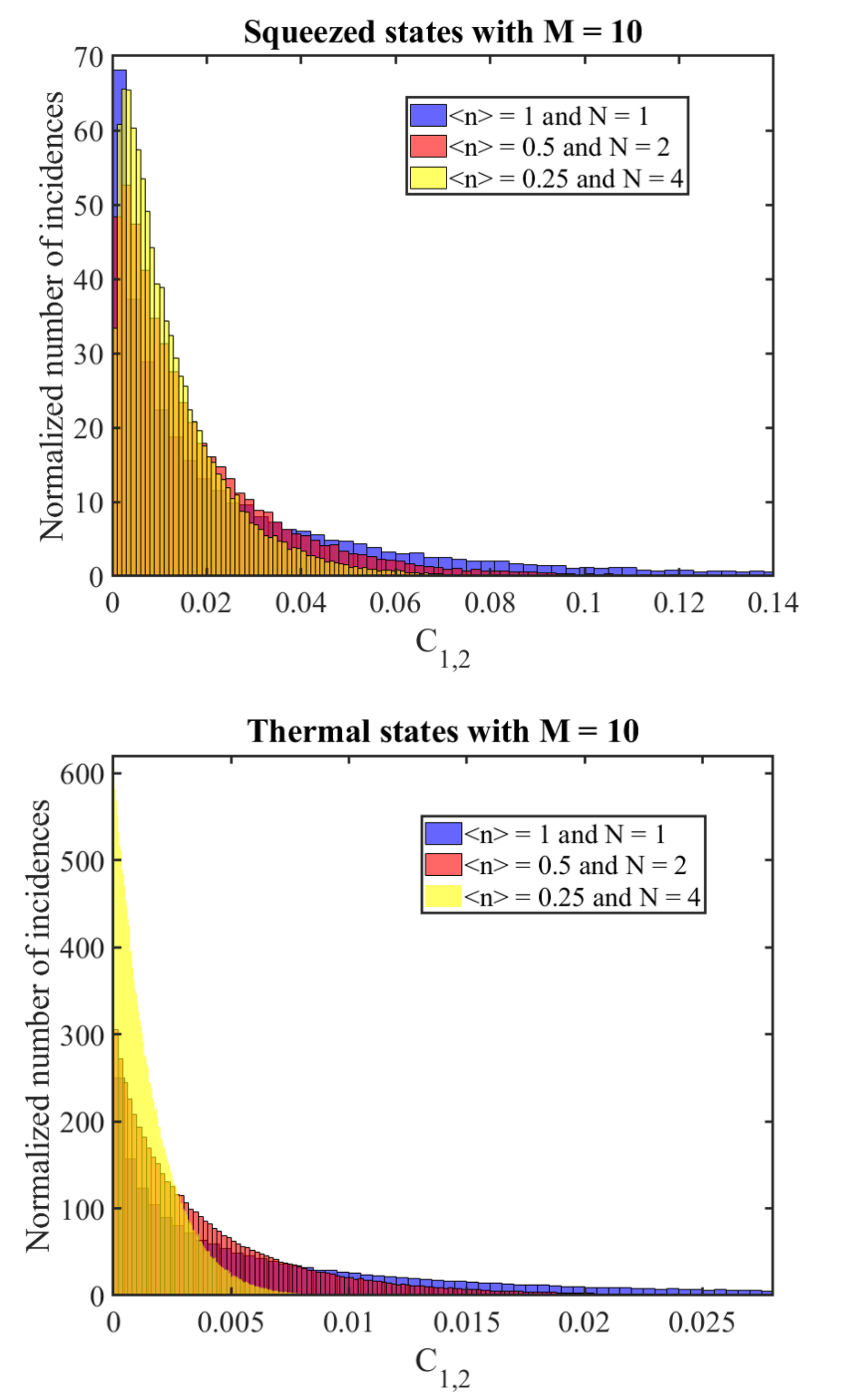}
	\caption{(Color online)
		The histograms of the two-point correlators $C_{1,2}$ for $\langle \hat{n}_\Sigma\rangle=1$, with $N=1$, $2$, and $4$.
		Squeezed (thermal) states are depicted in the top (bottom) plot.
	}\label{fig:ConstDilHists}
\end{figure}

	For this reason, we consider the mean total photon number $\langle \hat n_\Sigma\rangle$, where
	\begin{equation}
		\hat n_\Sigma=\sum_{j=1}^M \hat n_j.
	\end{equation}
	This mean value reads $\langle \hat n_\Sigma\rangle=N\langle \hat n\rangle$ for our scenario of $N$ occupied input modes with identical input states.
	For investigating the impact of the number of occupied modes, we keep the total energy constant while altering $N$, yielding $\langle\hat n\rangle=\langle\hat n_\Sigma\rangle/N$ for each nonvacuum input.
	Typical histograms for dilution can be seen in Fig. \ref{fig:ConstDilHists}.
	For a fair comparison, we additionally fix the number of modes $M$, regardless of the choice of occupied modes $N$.
	We make the particular choice of $M=10$ modes to satisfy the minimal constraint $M\gtrsim N^{2}$.

\begin{figure}[tb]
	\includegraphics[width=.9\columnwidth]{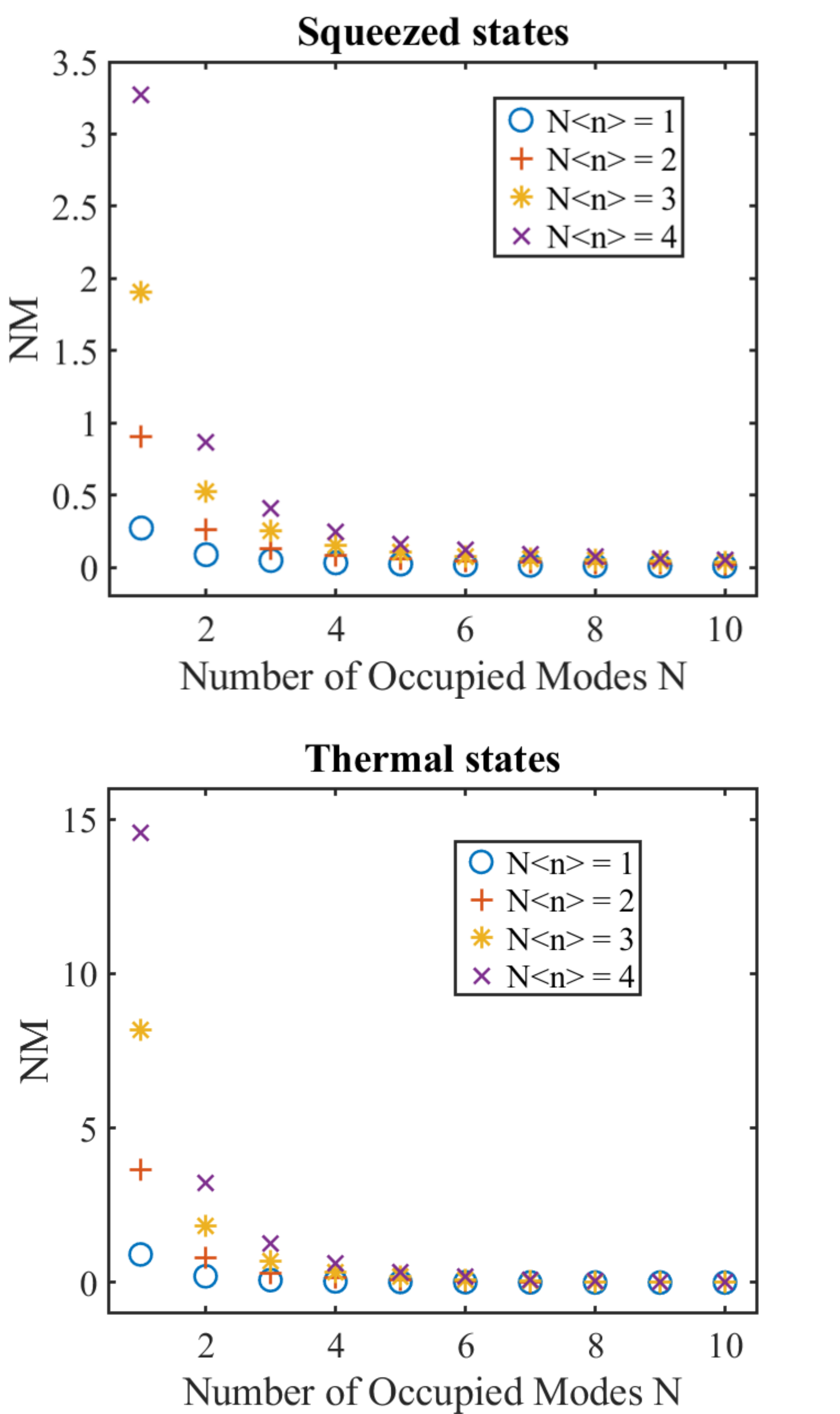}
	\caption{(Color online)
		Variation of $\mathsf{NM}$ with $N$ for $\langle \hat{n}_{\Sigma}\rangle\in\{1,2,3,4\}$.
		The normalized mean $\mathsf{NM}$ for squeezed (thermal) states is shown in the top (bottom) plot.
	}\label{fig:ConstDilNMs}
\end{figure}

	In Fig. \ref{fig:ConstDilNMs}, Eqs. \eqref{eq:RandomMatrixStatProp} and \eqref{eq:CMoments} are applied to plot the variation of $\mathsf{NM}$ for several values of $\langle \hat{n}_{\Sigma}\rangle$.
	We observe that the correlations are most pronounced for fewer occupied inputs with a higher mean photon number.
	This can be understood from the following considerations.
	For full dilution and spreading over all modes, $N=M$, the terms proportional to $\langle\hat n\rangle $ vanish in Eq. \eqref{eq:InOutSimpler}.
	Since these terms are always positive, their vanishing reduces the value of $C_{1,2}$.  
	Moreover, $\langle\hat\epsilon\rangle$ also takes its smallest value in the case of full dilution, such that this scenario must lead to the lowest $C_{1,2}$.
	If we treat unoccupied modes as asymmetries in the system, then larger asymmetries (i.e., small $N$ or large $\langle\hat n_\Sigma\rangle $) lead to larger two-point correlators.
	Thus, our method works best when remaining within the boson sampling limit (i.e., $M\gtrsim N^{2}$) in order to obtain stronger correlations.

\section{Experimental considerations}\label{sec:ExpCons}

	In a theoretical framework, one can assume that all states are created perfectly, all components are lossless, there is no noise, all unitaries are ideal, and each detector has a 100\% efficiency and an ideal photon-number resolution.
	In reality this is not the case.
	In this section, we therefore explore how the results from the previous sections are affected by such impurities and the tolerances required to obtain statistically significant results.

	For instance, we can think of experimental limitations in terms of state degradation, as measured, for example, by the state's purity.
	For a generic state density matrix $\hat{\rho}$, the state is pure if $\mathrm{Tr}\left(\hat{\rho}^{2}\right)=1$ and mixed if $\mathrm{Tr}\left(\hat{\rho}^{2}\right)<1$.
	For a Gaussian state, we can also invoke the relation
	\begin{equation}
	\label{eq:PurityRelation}
		\mathrm{Tr}\left(\hat{\rho}^{2}\right)=\frac{1}{\sqrt{\det\left(\boldsymbol{V}\right)}},
	\end{equation}
	which is true for any number of modes \cite{DMM94}.
	In general, the influence of a given imperfection onto the covariance matrix $\boldsymbol V$ determines the impact on the correlators.

	Further, imperfections can affect each individual mode in a different manner.
	This can be considered by using our general results from Sec. \ref{sec:GaussianCorr}.
	However, in practice, one can assume that all prepared states are subjected to almost the same amount of impurities when passing similar optical elements and being measured with similar detectors.
	Thus, for the sake of getting a fundamental idea of what the influence of different imperfections is, we can approximate imperfections by modeling them with identical influence on all modes.

\subsection{Network loss}

	The general description of multimode light propagating in a lossy network has been formulated, e.g., in Ref. \cite{GW96}.
	As outlined above, here we assume that the loss is homogeneously distributed.
	Thus, let $\eta$ be the overall quantum efficiency of the optical network and the detectors; i.e., the values $\eta=1$ and $\eta=0$ correspond to no loss and full loss, respectively.
	The well-known impact of loss on the characteristic quantities of Gaussian states reads
	\begin{equation}
	\label{eq:CovMatLoss}
		\boldsymbol{V}\mapsto
		\eta\boldsymbol{V}+(1-\eta)\boldsymbol{E}
		\quad\text{and}\quad
		\boldsymbol{\xi}\mapsto
		\sqrt{\eta}\boldsymbol{\xi},
	\end{equation}
	where $\boldsymbol E$ is the identity matrix.
	This means that the covariance matrix including loss is a convex mixture of the lossless covariance and the covariance matrix of the vacuum state (i.e., the identity matrix).
	In particular, the quadratures transform as $v_{s}\mapsto \eta v_{s}+(1-\eta)$, with $s\in\{q,p\}$.
	Thus, we get for the defining quantities of the correlator
	\begin{equation}
		\label{eq:LossChar}
		\langle\hat n\rangle
		\mapsto\eta\frac{v_{q}+v_{p}-2}{4}
		\quad\text{and}\quad
		\langle\hat \epsilon\rangle
		\mapsto\eta\frac{v_{q}-v_{p}}{4}.
	\end{equation}
	We can now use this to study the effect of loss on $C_{1,2}$ and its moments for a desired state, and the implications it has on distinguishing classical and quantum interference.	

	For example, we can consider a squeezed state at the input of a given mode.
	Using Eqs. \eqref{eq:SqueezeCov} and \eqref{eq:CovMatLoss}, the corresponding covariance matrix is then given by $\boldsymbol V=\eta\mathrm{diag}(e^{2r},e^{-2r})+(1-\eta)\mathrm{diag}(1,1)$.
	This means we obtain the purity from Eq. \eqref{eq:PurityRelation} as
	\begin{equation}
		\mathrm{Tr}\left(\hat{\rho}^{2}\right)=\left[4\eta\left(1-\eta\right)\sinh^{2}\left(r\right)+1\right]^{-1/2}.
	\end{equation}
	In addition, we can also characterize the purity through the uncertainty relation, which is minimally satisfied (i.e., $v_{q}v_{p}=1$) for pure Gaussian states.
	Including loss, we find for the squeezed state
	\begin{equation}
		\langle(\Delta \hat q)^{2}\rangle\langle(\Delta \hat p)^{2}\rangle
		=4\eta\left(1-\eta\right)\sinh^{2}\left(r\right)+1.
	\end{equation}
	Finally, from Eqs. \eqref{eq:InOutSimpler} and \eqref{eq:LossChar}, we can directly see that $C_{j,k}$ scales as
	\begin{equation}
	\label{eq:CjkLoss}
		C_{j,k}\mapsto\eta^{2}C_{j,k}.
	\end{equation}

\begin{figure*}
	\includegraphics[width=1\textwidth]{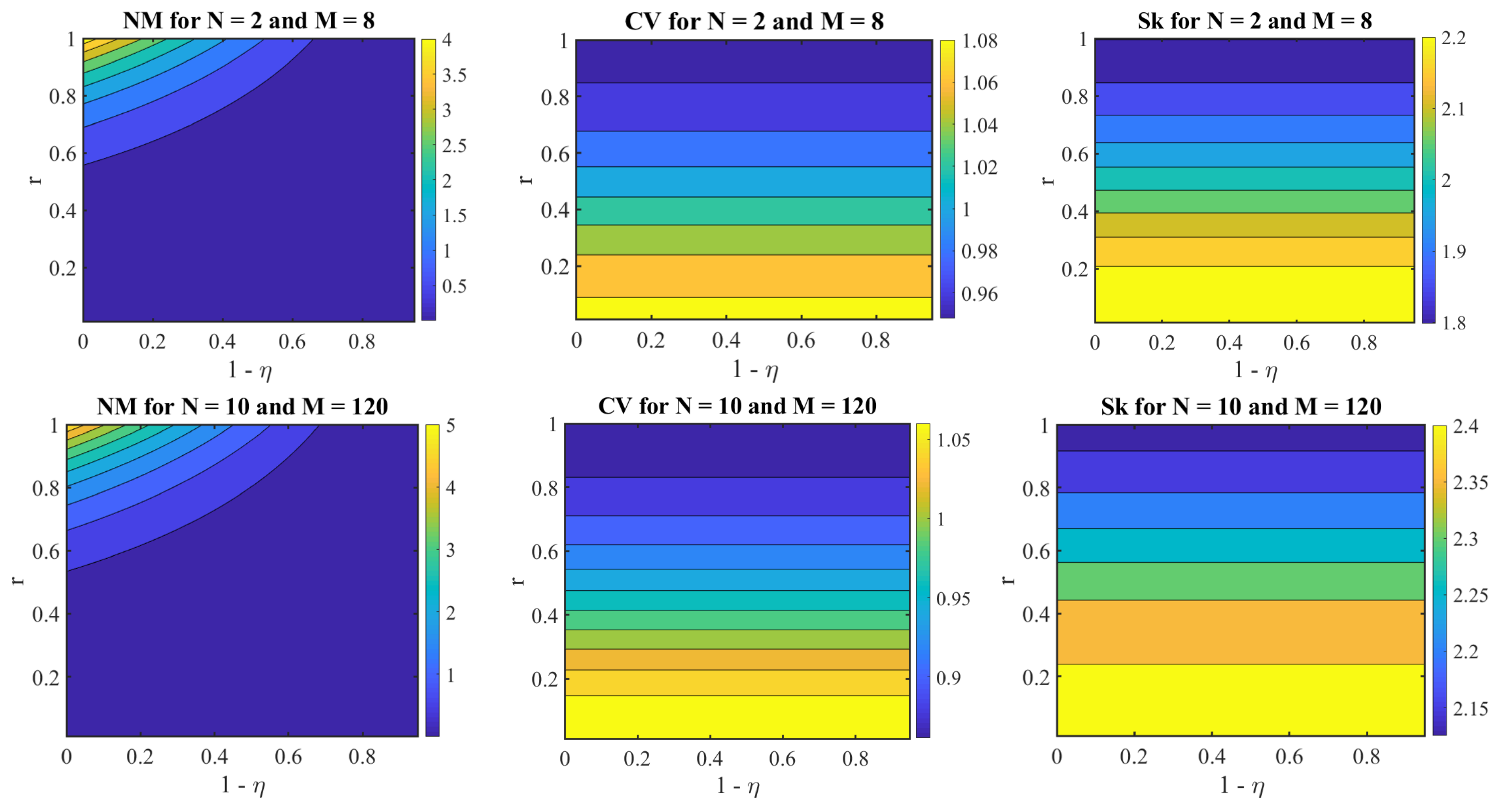}
	\caption{(Color online)
		Heat maps of $\mathsf{NM}$, $\mathsf{CV}$, and $\mathsf{Sk}$ (columns from left to right) for squeezed states with varying quantum efficiency $\eta$ and squeezing parameter $r$ for small ($N=2$ and $M=8$, top row) and large ($N=10$ and $M=120$, bottom row) optical networks.
		Note that $1-\eta$ is plotted on the horizontal axis for an increasing loss fraction.
	}\label{fig:SqueezeLossHeatmaps}
\end{figure*}

	The impact of loss [Eq. \eqref{eq:CjkLoss}] on $\mathsf{NM}$, $\mathsf{CV}$ and $\mathsf{Sk}$ for squeezed states is depicted in Fig. \ref{fig:SqueezeLossHeatmaps}.
	We see that $\mathsf{CV}$ and $\mathsf{Sk}$ do not vary with loss, which is intuitively clear from the definitions \eqref{eq:RandomMatrixStatProp} as the loss factors will cancel out.
	From this we might think that $\mathsf{CV}$ and $\mathsf{Sk}$ are good measures to tell squeezed states and thermal states apart.
	However, we will see in Sec. \ref{sec:DiscStatSig} that even with loss we get the most information with the least effort out of $\mathsf{NM}$.

\subsection{Additive noise}\label{AddingNoise}

	Adding Gaussian noise corresponds to a convolution with a Gaussian distribution.
	The thermal noise due to the environment is negligible for many optical settings.
	Still, other sources of noise have to be considered, for example, contributions from a nonideally filtered pump laser of the parametric process, etc.

	We take for our scenario $\boldsymbol V\mapsto\boldsymbol V+\boldsymbol V_\mathrm{noise}$, where the second term represents the convoluted noise contribution.
	Again, for simplicity, we assume that any quadrature for any mode is affected by the same noise, which gives $\boldsymbol V_\mathrm{noise}=\nu\boldsymbol E$.
	This leads to adapting the $v_{q}$ and $v_{p}$ parameters in mode $j$ as
	\begin{equation}
		v_{q}\mapsto v_{q}+\nu
		\quad\text{and}\quad
		v_{p}\mapsto v_{p}+\nu.
	\end{equation}
	Consequently, we find $\langle\hat n\rangle\mapsto \langle\hat n\rangle+\nu/2$, while the eccentricity remains unchanged, $\langle\hat\epsilon\rangle\mapsto\langle\hat\epsilon\rangle$.
	In addition, we arrive at
	\begin{equation}
		\langle(\Delta \hat q)^{2}\rangle\langle(\Delta \hat p)^{2}\rangle=1+\nu^{2}+2\nu\cosh\left(2r\right)
		=\frac{1}{[\mathrm{Tr}(\hat\rho^2)]^{2}}.
	\end{equation}
	From this we see that we only have a pure state when $\nu=0$, and any squeezing will exacerbate the purity.

	Another interesting property is considering subvacuum variances, i.e., squeezing.
	When the squeezing is along the $p$ quadrature, then we have $v_{p}=e^{-2r}\leq 1$ for a pure single-mode squeezed vacuum state ($r \geq 0$).
	It is interesting to consider the range of $r$ and $\nu$ for which squeezing is preserved, $v_{p}< 1$.
	We arrive at the condition
	\begin{equation}
	\label{eq:SubVacVar}
		r>-\frac{1}{2}\mathrm{ln}\left(1-\nu\right),
	\end{equation}
	which defines the boundary separating classical from squeezed states.

\begin{figure*}
	\includegraphics[width=1\textwidth]{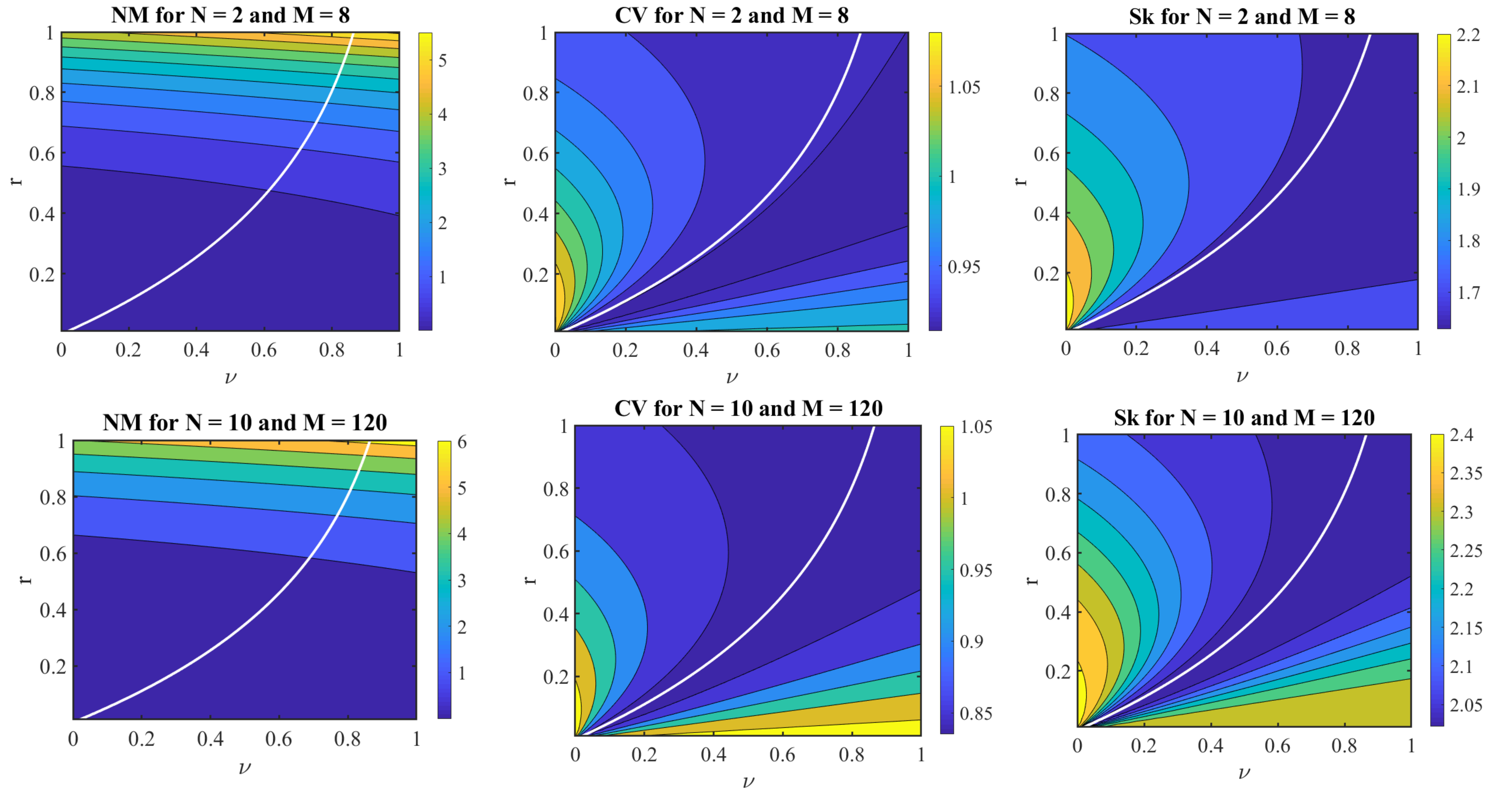}
	\caption{(Color online)
		Heat maps of $\mathsf{NM}$, $\mathsf{CV}$, and $\mathsf{Sk}$ (columns left to right) for noisy squeezed states obtained by varying squeezing parameter $r$ and noise parameter $\nu$, for a small ($N=2$ and $M=8$, top row) and large ($N=10$ and $M=120$, bottom row) systems.
		The (white) lines for the subvacuum variance boundary in Eq. \eqref{eq:SubVacVar} are additionally depicted; to the left of those lines, we have a subvacuum variance.
	}\label{fig:NoiseHeatmaps}
\end{figure*}

	In Fig. \ref{fig:NoiseHeatmaps}, we show the dependence of the values of $\mathsf{NM}$, $\mathsf{CV}$ and $\mathsf{Sk}$ on the noise and squeezing parameters.
	Compared to the loss scenario (cf. Fig. \ref{fig:SqueezeLossHeatmaps}), the functional landscape is more complex.
	Specifically, the coefficient of variation (center row) and the skewness (right row) exhibit nontrivial relations.
	Comparing the lower right-hand corner of the $\mathsf{CV}$ and $\mathsf{Sk}$ plots for both small (top row) and large (bottom row) systems in Fig. \ref{fig:NoiseHeatmaps}, it appears that the noise is suppressed by higher moments as the change with $\nu$ is more shallow in the $\mathsf{Sk}$ plots compared to the $\mathsf{CV}$ plots.

\subsection{Discrimination via statistical significance} \label{sec:DiscStatSig}

	In Secs. \ref{SqueezeThermalComp} and \ref{AddingNoise}, the discrimination of squeezed and thermal states using $\mathsf{NM}$, $\mathsf{CV}$ and $\mathsf{Sk}$ was discussed.
	It was suggested that $\mathsf{NM}$ would be suitable at higher average photon numbers and $\mathsf{Sk}$ at lower average photon numbers.
	However, when considering the experimental implications of this, the ideal situation is to use the metric which involves the least number of Haar-random unitaries, i.e., getting away with the fewest data points.

	In order to determine this, we approximate $\mathsf{NM}$, $\mathsf{CV}$ and $\mathsf{Sk}$ by generating a set of $C_{1,2}$ from several Haar-random unitaries and using the relations in Eq. \eqref{eq:RandomMatrixStatProp}.
	The variation in values of $C_{1,2}$ enables us to assign statistical uncertainties to those quantities, using the typical propagation of errors.
	Further, we consider $\Delta \mathsf{NM}=\mathsf{NM}_{S}-\mathsf{NM}_{T}$ (and equivalent for $\mathsf{CV}$ and $\mathsf{Sk}$), which is the difference of this the normalized mean for squeezed and thermal states.
	The metric we impose is the minimum number of iterations required for $\Delta \mathsf{NM}$ to be nonzero with a $3\sigma$ statistical significance.
	This statistical bound is enough to tell the considered families of states apart.
	Specifically, we find that the minimal number is not massively affected by loss or system size, which is discussed in the following.

\begin{figure}[tb]
	\includegraphics[width=1\columnwidth]{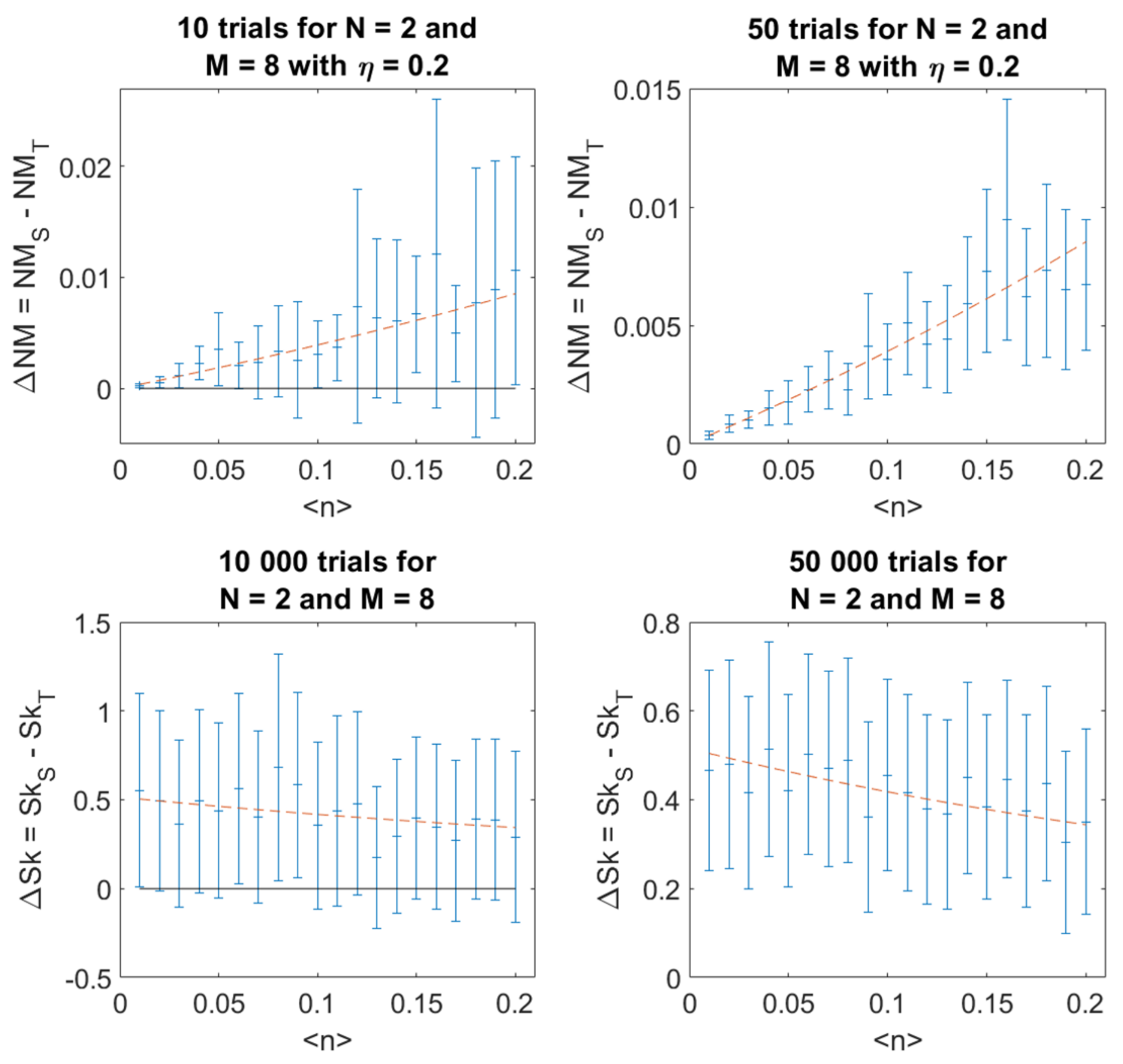}
	\caption{(Color online)
		The top row shows $\Delta \mathsf{NM}$ with $3\sigma$ error bars plotted against average photon number $\langle \hat{n}\rangle $ for small systems ($N=2$ in $M=8$) with $\eta=0.2$ (i.e., $80\%$ loss).
			On the left, the system evolved under $10$ different Haar-random unitaries (i.e., $10$ trials); we considered $50$ trials for the plot on the right.
		The bottom row depicts $\Delta \mathsf{Sk}$ with $3\sigma$ error bars plotted against average photon number $\langle \hat{n}\rangle $ for small systems ($N=2$ in $M=8$).
			The left and right plots use $10\,000$ and $50\,000$ trials, respectively.
		The blue error bars are from the simulated data, and orange dotted lines are from the analytical expressions.
		Black zero lines have been added to the plots in the left column to show that the error bars go through zero.
	}\label{fig:NMStatDisc}\label{fig:SkStatDisc}
\end{figure}

	From our previous analysis, we can see that $\Delta \mathsf{NM}$ scales with $\langle \hat{n}\rangle $, so only low values of $\langle \hat{n}\rangle $ are considered at the top of Fig. \ref{fig:NMStatDisc}.
	The top two plots contain parameters typical for current photonic architectures.
	We can see that 10 trials are not enough, but 50 trials discriminate between squeezed and thermal states with a $3\sigma$ significance.
	In addition to this, further simulations showed that the degree of system loss does not affect the number of trials required for discrimination.
	This result is encouraging, as 50 trials seems to be a feasible number to undertake in a laboratory.

	It was proposed in Sec. \ref{SqueezeThermalComp} that $\mathsf{Sk}$ might be a good measure to discriminate between squeezed and thermal states at low values of $\langle \hat{n}\rangle $ by considering Fig. \ref{fig:NMCVSk}.
	Therefore, we consider the same for $\Delta \mathsf{Sk}$ with varying $\langle \hat{n}\rangle $.
	The results can be found at the bottom of Fig. \ref{fig:SkStatDisc} for small systems ($N=2$ in $M=8$); losses were not considered as loss does not effect $\mathsf{Sk}$ [cf. Fig. \ref{fig:SqueezeLossHeatmaps}].
	We see that $10\,000$ trials are insufficient for a discrimination, but $50\,000$ allow this with a statistical significance of $3\sigma$.
	The meaning of this result is that on the order of $10^{3}$ more trials are required when using $\mathsf{Sk}$ compared to $\mathsf{NM}$ to tell squeezed and thermal states apart, regardless of the value of $\langle \hat{n}\rangle $.
	Therefore, $\mathsf{NM}$ is clearly the best metric as it works sufficiently well, even in the presence of system loss.

\subsection{Detectors with finite photon-number resolution}\label{sec:DetectorPNR}

	It was mentioned in Sec. \ref{SimMeth} that one of the simulation methods involved projecting the output state onto the photon-number basis and calculating $C_{1,2}$ from the obtained statistics.
	If $P\left(n_{1},n_{2}\right)$ is the probability of detecting $n_{1}$ photons in mode 1 and $n_{2}$ photons in mode 2, then $C_{1,2}$ is given by
	\begin{equation}
	\label{eq:C12Exp}
	\begin{aligned}
		C_{1,2}
		=&\sum_{n_1,n_2=0}^\infty n_{1}n_{2}P\left(n_{1},n_{2}\right)
		\\
		&-\left(\sum_{n_1=0}^\infty n_{1}P\left(n_{1}\right)\right)\left(\sum_{n_2=0}^\infty n_{2}P\left(n_{2}\right)\right),
	\end{aligned}
	\end{equation}
	where the marginal distributions are given by $P\left(n_{1}\right)=\sum_{n_2=0}^\infty P\left(n_{1},n_{2}\right)$ and $P\left(n_{2}\right)=\sum_{n_1=0}^\infty P\left(n_{1},n_{2}\right)$.

	The output of the simulation after tracing over all but modes 1 and 2 is a matrix where the entries correspond to $P\left(n_{1},n_{2}\right)$.
	Equations \eqref{eq:SqueezeFock} and \eqref{eq:ThermalFock} yield that the contributions for high photon numbers become arbitrarily small for both squeezed and thermal states.
	Therefore, for a finite simulation, we can consider a highest sensible photon number and truncate our statistics without affecting the result.

	In fact, such a truncation resembles a common experimental restriction to photon-number detectors.
	It is possible to multiplex detectors with a finite maximal photon-number resolution \cite{HBLNGS16,Setal17}, such as transition edge sensors (TESs), to increase the maximally measurable photon number.
	Still, each TES is restricted to about 11 photons, which consequently poses a significant limitation to the mean photon number for an experiment.
	Therefore, the maximum photon-number resolution the detectors are capable of is an important consideration.

	It is also worth mentioning that recent developments in Gaussian boson sampling theory have lead to extending the framework to click detectors \cite{QAK18}, where it is shown that the problem has the same unfavorable scaling for low squeezing.
	However, as we will see, photon number resolution is still required for a measurement of $C_{1,2}$ with low error.

	In order to test this, for a given system evolved under a given Haar-random unitary, $P\left(n_{1},n_{2}\right)$ can be calculated up to a level much higher than a TES is capable of (here, for up to 40 photons per mode).
	Then this sample enables us to approximate $C_{1,2}$ via Eq. \eqref{eq:C12Exp} by truncating at successively higher values of maximal photon numbers ($n_{1},n_{2}\leq n_{\max}$).
	These values are then compared to the analytical value for the same system and Haar-random unitary using Eq. \eqref{eq:InOutSimpler}, and the relative distance $[C_{1,2}^\mathrm{(analytical)}-C_{1,2}^\mathrm{(estimated)}]/C_{1,2}^\mathrm{(analytical)}$ between the exact and estimated result can be calculated.

\begin{figure}[tb]
	\includegraphics[width=1\columnwidth]{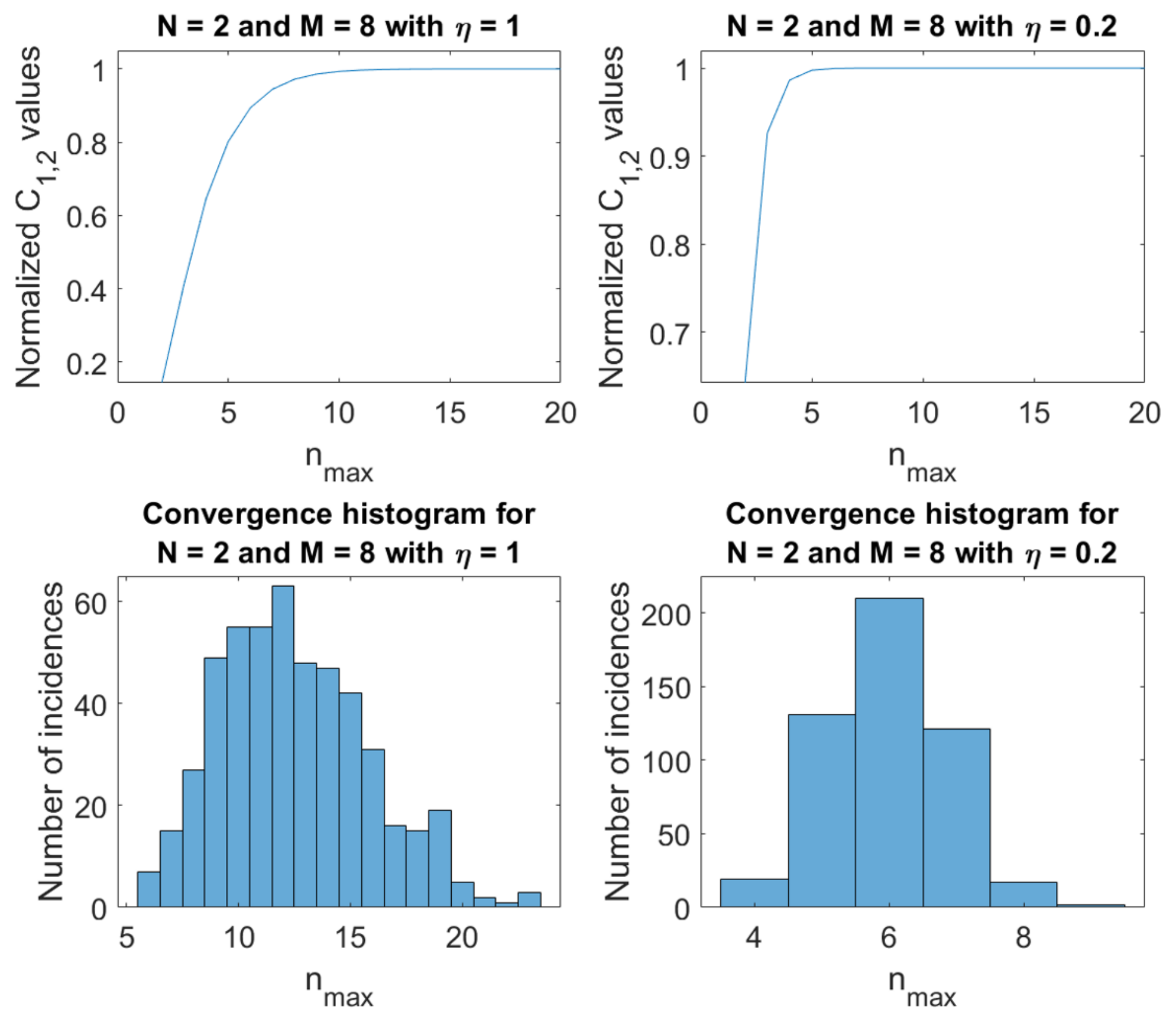}
	\caption{(Color online)
		The top row depicts convergence plots of $C_{1,2}$ for an example Haar-random unitary (which have been normalized against the exact value) plotted against maximum photon number resolution  $n_{\max}$.
			Both consider a small system ($N=2$ in $M=8$) with $\langle \hat{n}\rangle=1$.
			On the left, we have a system with full transmission $\eta=1$.
			On the right, we consider a system with $\eta=0.2$, typical of current architectures.
		The bottom plot shows two histograms for the number of incidences of $n_{\max}$ that satisfy convergence condition in Eq. \eqref{eq:NumResErrorBound} for 500 different Haar-random unitaries ($\eta=1$ on the left and $\eta=0.2$ on the right).
	}\label{fig:C12ErrorBound}
\end{figure}

	Examples for the desired convergence with the maximal photon number $n_{\max}$ can be seen at the top of Fig. \ref{fig:C12ErrorBound}.
	For statistical analysis, it is useful to say that suitable relative distance of smaller than $10^{-3}$ should be achieved, i.e.,
	\begin{equation}
	\label{eq:NumResErrorBound}
		-\log_{10}\left(\frac{C_{1,2}^\mathrm{(analytical)}-C_{1,2}^\mathrm{(estimated)}}{C_{1,2}^\mathrm{(analytical)}}\right)>3.
	\end{equation}
	The required value of $n_{\max}$ to achieve this will depend on the system in question.
	A lossy system will have lower $\langle \hat{n}\rangle $ on average compared to a lossless system, so a lower resolution would be required for the same convergence.
	Also, we observe that for $N=10$ occupied modes in an $M=120$-mode system, we have a more dilute photon number distribution at the output compared to $N=2$ and $M=8$.
	Thus, the former case also requires a lower resolution.
	As the measurement should be done for different unitaries, typical histograms can be additionally seen at the bottom of Fig. \ref{fig:C12ErrorBound}.
	The spread in values arises due to the variation in scattering of different unitaries to the output ports in consideration.
	Even with only 500 different Haar-random unitaries, it shows there is a mean resolution for the convergence condition in Eq. \eqref{eq:NumResErrorBound}.
	Interestingly, squeezed state inputs require a slightly higher resolution compared to a thermal state with the same mean photon number.
	This is specifically due to the fact that higher photon-number correlations scale differently for these classes of states, even though the mean photon number is the same.
	Moreover, typical experimental parameters for current architecture would be two single-mode squeezed vacuum inputs with $\langle \hat{n}\rangle \approx1$ each with $80\%$ loss per mode, which corresponds to the right column in Fig. \ref{fig:C12ErrorBound}.
	Therefore, a TES resolution ($n_{\max}=11$) would be enough to measure $C_{1,2}$ to within the error bound.
	If the loss were reduced, the squeezing could be even further reduced to lower $\langle \hat{n}\rangle $, allowing for a reduced $n_{\max}$.

\section{Summary and conclusions}

	In summary, we established methods for benchmarking boson sampling in realistic setups with Gaussian input states.
	Based on a previously introduced technique \cite{WKUMTRB16} applicable to phase-insensitive Fock states, we derived an analytical expressions for the intensity correlation between pairs of output detectors.
	In particular, these correlations are found to be affected by the eccentricity of the initial states' uncertainty ellipses.
	This effect is not present in the standard boson sampling setup.
	The corresponding additional terms in the correlators may indicate a previously unstudied type of many-particle interference phenomenon in this setting that is induced through squeezing.
	The resulting different structure of the two-point correlators translates to a quantitative difference upon averaging over all possible unitary circuits.
	By virtue of random matrix theory, these averages could then be evaluated analytically, which provides us with a predictive tool to recognize faulty Gaussian boson samplers.

	Furthermore, our results enable us to efficiently distinguish nonclassical squeezed vacuum states from classical thermal input states.
	This is an important finding as sampling from the latter states can be simulated efficiently with classical resources, while this is not the case for the former states.	
	In addition, we observed a clear difference in the properties of the correlations when few modes with highly squeezed input states are compared to many modes with weakly squeezed input states for a constant expectation value of the total particle number.

	We then employed the obtained properties of the two-point correlators as a tool to assess experimental constraints.
	In the standard boson sampling setup, losses can be eliminated through post-selection, even though this has a negative effect on the sampling efficiency because of a decreased number of accounted events, and therefore on the reasonableness of any claim to ``quantum advantage."
	For Gaussian boson sampling, loss and noise processes must be taken into account explicitly; we were able to perform this task when applying our general approach.
	Typically, these imperfections have the advantage of being Gaussian such that they can be simply incorporated in the initial state.
	We identified the average two-point correlation as a good robust certifier of Gaussian boson sampling, even in the presence of attenuation and other noise processes.
	Additionally, we showed that the rescaled higher moment---the coefficient of variation and the skewness---are unaffected by loss.
	However, these higher moments do show interesting features in the presence of classical Gaussian noise.
	In particular, we show that the second and third moments can be used as probes for transitions from nonclassical to classical light, which occurs when the classical noise drives the quadrature fluctuations beyond the shot-noise level.
	Ultimately, we find that the mean value for the two-point correlators is the most useful quantity at our disposal since it can be obtained with rather low statistical fluctuations from relatively low sample sizes.
	The higher moments, as represented by the coefficient of variation and the skewness, require much more effort to reach convergence in the statistics.

	Finally, we explored the feasibility of performing the proposed correlation tests for Gaussian boson sampling experimentally with state-of-the-art photon-number resolving detectors.
	These results suggest that for small photonic circuits with a small number of occupied input ports, we may only use a subset of possible random photonic circuits.
	As the number of modes increases and we consider larger circuits, we observe that the requirements on the level of photon counting become less stringent.
	Ultimately, this implies that the presented methods are well suited for implementations in large-scale boson sampling setups.

	Here we compared interesting classes of phase-sensitive quantum states from a fundamental physics perspective, and, moreover, explored the impact of several important error models.
	In essence, we found that often the mean two-point correlation is already sufficient to distinguish different classes of inputs.
	It remains an open question whether there is a genuinely challenging attack for Gaussian boson sampling, as was the case with the mean field sampler in the standard boson sampling scenario \cite{WKUMTRB16}.
	Similarly, it is an intriguing open question in what sense the present results might be affected by the possibility of having (partial) distinguishability in the additional degrees of freedom of the input states, e.g., through different polarizations or time-frequency modes.

\begin{acknowledgments}
	The authors would like to thank Ra\'ul Garc\'{\i}a-Patr\'on for enlightening discussions.
	This work has received funding from the European Union's Horizon 2020 Research and Innovation Program under Grant Agreement No. 665148 (QCUMbER).
	D.S.P. acknowledges funding through the Networked Quantum Information Technologies (NQIT) hub (part of the UK National Quantum Technologies Programme) under Grant No. EP/N509711/1.
	M.W. acknowledges funding through research fellowship WA 3969/2-1 from the German Research Foundation (DFG).
	J.J.R. acknowledges funding through NWO Rubicon.
\end{acknowledgments}

\appendix*
\section{Moments of Gaussian states}\label{app:Moments}

	A convenient method to access the moments of a distribution is formulated in terms of characteristic functions, the Fourier transform of the initial distribution.
	In Ref. \cite{VW06}, a comprehensive introduction to characteristic functions for quantum-optical phase-space distributions can be found.
	Here, let us recall some concepts which are essential for our purposes.

	The characteristic function to the Glauber-Sudarshan distribution is the normally ordered expectation value of the displacement operator, taking the form
	\begin{equation}
		\Phi(\boldsymbol \beta)=\langle
			e^{\beta_1\hat a_1^\dag}e^{-\beta_1^\ast\hat a_1}
			\cdots
			e^{\beta_M\hat a_M^\dag}e^{-\beta_M^\ast\hat a_M}
		\rangle
	\end{equation}
	for an $M$-mode quantum state of light and the complex vector $\boldsymbol\beta=(\beta_1,\ldots,\beta_M)^\mathrm{T}$.
	The characteristic function satisfies $\Phi(0)=1$ (normalization) and $\Phi(-\boldsymbol \beta)=\Phi(\boldsymbol \beta)^\ast$ (Hermiticity).
	The derivatives of the characteristic function relate to the moments of the distribution,
	\begin{equation}
	\begin{aligned}
		&\partial_{\beta_1}^{j_1}\cdots\partial_{\beta_M}^{j_M}
		\partial_{\beta_1^\ast}^{k_1}\cdots\partial_{\beta_M^\ast}^{k_M}
		\Phi(\boldsymbol\beta)|_{\boldsymbol\beta=0}
		\\
		=&(-1)^{k_1+\cdots+k_M}\langle
			\hat a_1^{\dag j_1}\cdots\hat a_M^{\dag j_M}
			\hat a_1^{k_1}\cdots\hat a_M^{k_M}
		\rangle.
	\end{aligned}
	\end{equation}
	Furthermore, the reformulation $\hat a_l=\Delta\hat a_l+\alpha_{0,l}$ results in the characteristic function
	\begin{equation}
		\Phi_{\Delta}(\boldsymbol \beta)=e^{\boldsymbol\beta^\dag\boldsymbol \alpha_0-\boldsymbol\alpha_0^\dag\boldsymbol \beta}\Phi(\boldsymbol \beta)
	\end{equation}
	for central moments.

	For our purposes, we are specifically interested in Gaussian states.
	In this case, the characteristic function is known to simplify to
	\begin{equation}
		\Phi_{\Delta}(\boldsymbol\beta)=e^{\vartheta(\boldsymbol\beta)},
	\end{equation}
	where we used the second-order polynomial
	\begin{eqnarray}
		\nonumber
		\vartheta(\boldsymbol\beta)
		&=&\frac{1}{2}\sum_{j,k=1}^M\Big(
			\langle \Delta\hat a_j^\dag\Delta\hat a_k^\dag\rangle\beta_{j}\beta_{k}
			+\langle \Delta\hat a_j\Delta\hat a_k\rangle\beta_{j}^\ast\beta_{k}^\ast
		\Big)
		\\
		&&-\sum_{j,k=1}^M\langle \Delta\hat a_j^\dag\Delta\hat a_k\rangle\beta_{j}\beta_{k}^\ast.
	\end{eqnarray}
	It is worth recalling that all except the second-order derivatives of $\vartheta$ vanish for $\boldsymbol\beta=0$.
	From the derivatives of this specific characteristic function, we find the following relation for the central fourth-order moments for Gaussian states:
	\begin{equation}
	\begin{aligned}
		&\langle
		\partial_{\beta_r}\partial_{\beta_s}\partial_{\beta_t^\ast}\partial_{\beta_u^\ast}\Phi_{\Delta}(\boldsymbol\beta)|_{\boldsymbol \beta}
		=\langle
			\Delta\hat a_r^\dag\Delta\hat a_s^\dag\Delta\hat a_t\Delta\hat a_u
		\rangle
		\\
		=&\langle
			\Delta\hat a_r^\dag\Delta\hat a_s^\dag
		\rangle\langle
			\Delta\hat a_t\Delta\hat a_u
		\rangle
		{+}\langle
			\Delta\hat a_r^\dag\Delta\hat a_t
		\rangle\langle
			\Delta\hat a_s^\dag\Delta\hat a_u
		\rangle
		\\
		&{+}\langle
			\Delta\hat a_r^\dag\Delta\hat a_u
		\rangle\langle
			\Delta\hat a_s^\dag\Delta\hat a_t
		\rangle.
	\end{aligned}
	\end{equation}
	Using the bosonic commutation relations for $\hat a_s^\dag$ and $\hat a_t$, we can express the sought-after moments in terms of normally ordered moments.
	This finally yields
	\begin{equation}
	\begin{aligned}
		\langle
			\Delta\hat a_r^\dag\Delta\hat a_t\Delta\hat a_s^\dag\Delta\hat a_u
		\rangle
		=&\langle
			\Delta\hat a_r^\dag\Delta\hat a_s^\dag
		\rangle\langle
			\Delta\hat a_t\Delta\hat a_u
		\rangle
		\\
		&{+}\langle
			\Delta\hat a_r^\dag\Delta\hat a_t
		\rangle\langle
			\Delta\hat a_s^\dag\Delta\hat a_u
		\rangle
		\\
		&{+}\langle
			\Delta\hat a_r^\dag\Delta\hat a_u
		\rangle\langle
			\Delta\hat a_t\Delta\hat a_s^\dag
		\rangle.
	\end{aligned}
	\end{equation}


\end{document}